\documentclass[12pt]{article}

\usepackage[utf8]{inputenc}
\usepackage{authblk}
\usepackage{amsmath, amssymb}
\usepackage[T1]{fontenc}
\usepackage{wrapfig}
\usepackage{float}
\usepackage[breaklinks]{hyperref}
\usepackage{color}
\usepackage{graphicx}
\hypersetup{colorlinks=true, linkcolor=black, citecolor=black, filecolor=blue, urlcolor=blue, unicode=true}
\usepackage{longtable}
\usepackage{booktabs}
\usepackage{subcaption}
\usepackage{comment}
\usepackage{multirow,bigdelim}
\usepackage{calc}
\usepackage[authoryear]{natbib}
\newcommand{\doa}{September 15, 2020}

\begin{document}

\title{Regional now- and forecasting for data reported with delay:
	Towards surveillance of COVID-19 infections}
\author[*]{Giacomo De Nicola}
\author[*]{Marc Schneble}
\author[*]{G{\"o}ran Kauermann}
\author[**]{Ursula Berger}

\affil[*]{Department of Statistics, Ludwig-Maximilians-University
  Munich, Germany}
\affil[**]{ Institute for Medical Information Processing, Biometry and Epidemiology,
Ludwig-Maximilians-University
  Munich, Germany}

\renewcommand\Authands{ and }
\date{\vspace{-5ex}}

\maketitle

\begin{abstract}
Governments around the world continue to act to contain and mitigate the spread of COVID-19. The rapidly evolving situation compels officials and executives to continuously adapt policies and social distancing measures depending on the current state of the spread of the disease. In this context, it is crucial for policymakers to have a firm grasp on what the current state of the pandemic is as well as to have an idea of how the infective situation is going to unfold in the next days. However, as in many other situations of compulsorily-notifiable diseases and beyond, cases are reported with delay to a central register, with this delay deferring an up-to-date view of the state of things. We provide a stable tool for monitoring current infection levels as well as predicting infection numbers in the immediate future at the regional level. We accomplish this through nowcasting of cases that have not yet been reported as well as through predictions of future infections. %The two steps are also combined in forenowcasting.
 We apply our model to German data, for which our focus lies in predicting and explain infectious behavior by district.    
\end{abstract}

\section{Introduction}
The infectious disease known as COVID-19 hit the planet in tsunami-like fashion. The first cases were identified in December 2019 in the city of Wuhan, China, and by March 2020 infections had already spread over the entire world. Nearly all of the affected countries progressively implemented measures to slow down the spread of the virus, ranging from recommended social distancing to almost complete lockdowns of social and economic activity. These measures eventually proved to be effective, as the number of infections could be slowed down (see e.g. \citealp{Flaxman2020} and \citealp{Roux2020}). This allowed numerous states to relax restrictions, in an attempt to gradually return to normality. At the same time, with the threat posed by the virus still looming, decision makers are forced to strike a balance between epidemiological risk and allowance  of socio-economic activity. In this context, surveillance of the number of new infections became increasingly important, and particularly so on a regional level. Given the local nature of the phenomenon (see e.g. \citealp{Gatto2020} and \citealp{Li489}), such regional view appears to be of crucial importance. One of the difficulties lies in the fact that exact numbers of infections detected on a particular day are only available with a reporting delay of, in some cases, several days, which occurs along the reporting line from local health authorities to the central register. The following paper provides a stable tool for monitoring current infection levels corrected for  incompleteness of the data due to reporting delays. This approach is also extended towards predicting the infectious behavior in the immediate future at the regional level. %After formulating it, we apply our model to explain and predict numbers of registered COVID-19 infections for Germany by district.

More specifically, the scope of our model is threefold: Firstly, we aim to understand the current epidemiological situation as well as to comprehend the association between the number of detected infections and demographic characteristics and geographical location. Secondly, our goal is to nowcast infections that have already been observed but have not yet been included in the official numbers. New infections are detected through tests and registered by the local health authorities, which in turn will report the numbers to national authorities with an inevitable delay. Since we observe reports of infections for each day, we are able to model this delay, which indeed allows to nowcast infection numbers for today correcting for infections which have not yet been reported. Lastly, our aim is also to forecast the epidemiological situation for the immediate future. We here want to stress that our model is not aiming to exactly predict future infection numbers, as that would not be realistic. The goal is rather to try and give a general idea of what is going to happen in the next days in the different districts, and, perhaps most importantly, help identify which districts are going to be the most problematic. This could also help policymakers in making decisions regarding the implementation of safety measures at the regional level. 

We apply our modeling approach to explain and predict numbers of registered COVID-19 infections for Germany by district, age group and gender. While the regional component is of evident and paramount importance, the age group and gender distinctions are also very relevant, given the powerful interaction of demography and current age-specific mortality for COVID-19 \citep{Dowd9696}.

Our nowcasting approach can also be used to obtain up-to-date measures of the 7-days incidence, both at the local as well as at the national level. This quantity is often used by authorities to assess how hard a specific area is currently hit by the pandemic, and sometimes (as is the case for Germany) it is also employed as a criterion to decide which containment measures are appropriate. It is especially important to have up-to-date infection numbers when computing such a measure, as it is inherently evolving on a daily basis; At the time of writing, the index is calculated by German officials through use of the date of report by the local health authorities. Given that, as already stated, there are significant delays in the reporting of cases from local authorities to the national ones, the resulting figures are consistently underestimating the actual incidence, with the error being potentially quite large and problematic. Our nowcasts offer a simple and stable solution to this issue, providing infection numbers that are already corrected for expected delays. 

The statistical modeling of infectious diseases is a well developed scientific field. We refer to \citet{Held-etal:2017} for a general overview of the different models. Modeling and forecasting COVID-19 infections has been tackled by numerous research groups using different models.  \citet{Panovska:2020} discusses whether one or multiple models may be useful for COVID-19 data analytics. \citet{Stuebinger:2020} make use of time warping to forecast COVID-19 infections for different countries (see also \citealp{Cintra:2020}), while \citet{Dehesh:2020} utilize ARIMA time series models. \cite{Ray2020} combine forecasts from several different models to obtain robust short-term forecasts for deaths related to COVID-19. \cite{Fritz2021} present a multimodal learning approach combining statistical regression and machine learning models for predicting COVID-19 cases in Germany at the local level. Early references dating back to the first stages of the pandemic are   \citet{Anastassopoulou:2020} and  \citet{Petropoulos:2020}. In this paper 
% We apply our model to Germany, and since our focus is on analysing the district-specific dynamics of fatal infections, we follow an approach similar to that used by \cite{schneble-etal:2020} for the nowcasting of COVID-19 related deaths.  More specifically, 
we make use of negative binomial regression models implemented in the {\tt mgcv} package in \texttt{R} \citep{Wood:2017}. This allows us to decompose the spatial component in depth, and obtain district-level nowcasts and forecasts for Germany. %One of our intentions is to demonstrate that today's "off the shelf" software can cope with the problem in an efficient as well as flexible way once the modeling problem has been defined in a suitable form. 
Our results confirm the dynamic and highly local nature of outbreaks, highlighting the need for continuous regional surveillance on a small area level.

The rest of the paper is structured as follows: Section \ref{sec:data} describes the data, while Section \ref{sec:surveillance} frames the problem, presents our model and compares the performance of different variants over time, motivating our modeling choices. Section \ref{sec:predictions} exemplifies surveillance and describes how predictions are performed in practice, showing the results for exemplary dates. Finally, Section \ref{sec:discussion} concludes the paper, highlighting the limitations of this study and adding some concluding remarks.

\section{Data}
\label{sec:data}

As previously anticipated, we focus our analyses on German data. To do so, we make use of the COVID-19 dataset published by
the Robert-Koch-Institute (RKI) on a daily basis. 
The RKI is a German federal government agency and scientific institute responsible for health reporting and for disease control and prevention. It maintains the national register for COVID-19, where all identified cases of this compulsorily-notifiable disease are reported from the local health authorities to the RKI.  
In our analysis we make use of daily downloads of the data, which we have at our disposal starting from April 12, 2020 until December 29, 2020. As we want our model to be dynamic, the surveillance analysis is performed considering only infections with registration dates within 21 days of the day of analysis, with earlier data allowing us to compare the differences in model fit over time. 

\begin{table}
	\center
	\resizebox{\textwidth}{!}{\begin{tabular}{lcccccc}
			\toprule
			\ldelim\{{7}{*}[\parbox{3.6cm-\tabcolsep-\widthof{$\Bigg]$}}{
				Data downloaded on September 25, 2020}] & District & Age Group & Gender & Infections  & Registration Date & Reporting Date \\
			& (Landkreis) & (Altersgruppe) & (Geschlecht) & (Anzahl Fall) & (Meldedatum) & (Datenstand) \\
			\cmidrule{2-7}
			& \multicolumn{1}{c}{$\vdots$} & \multicolumn{1}{c}{$\vdots$}& \multicolumn{1}{c}{$\vdots$}&  \multicolumn{1}{c}{$\vdots$}& \multicolumn{1}{c}{$\vdots$}& \multicolumn{1}{c}{$\vdots$} \\
			& Munich City & 60-79 & F & 3  & September 22, 2020 & September 25, 2020 \\
			& Munich City & 60-79 & M & 5 & September 22, 2020 & September 25, 2020 \\
			&  \multicolumn{1}{c}{$\vdots$} & \multicolumn{1}{c}{$\vdots$}& \multicolumn{1}{c}{$\vdots$}& \multicolumn{1}{c}{$\vdots$}&  \multicolumn{1}{c}{$\vdots$}& \multicolumn{1}{c}{$\vdots$} \\
			\midrule
			\ldelim\{{6}{*}[\parbox{3.6cm-\tabcolsep-\widthof{$\Bigg]$}}{
				Data downloaded on September 26, 2020}] &  District & Age Group & Gender & Infections  & Registration Date & Reporting Date \\
			\cmidrule{2-7}
			&  \multicolumn{1}{c}{$\vdots$} & \multicolumn{1}{c}{$\vdots$}& \multicolumn{1}{c}{$\vdots$}&  \multicolumn{1}{c}{$\vdots$}&  \multicolumn{1}{c}{$\vdots$}& \multicolumn{1}{c}{$\vdots$} \\
			&  Munich City & 60-79 & F & 6 & September 22, 2020 & September 26, 2020 \\
			& Munich City & 60-79 & M & 5 & September 22, 2020 & September 26, 2020 \\
			&  \multicolumn{1}{c}{$\vdots$} & \multicolumn{1}{c}{$\vdots$}& \multicolumn{1}{c}{$\vdots$}&  \multicolumn{1}{c}{$\vdots$}& \multicolumn{1}{c}{$\vdots$}& \multicolumn{1}{c}{$\vdots$} \\
			\bottomrule
	\end{tabular}}
	\caption{Illustration of the raw data structure, showing downloads of the data from September 25 and September 26, 2020 as an example. To facilitate reproducibility, the original column names used in the RKI datasets are given in brackets below our English notation.}
	\label{tab:data_structure}
\end{table}

%\begin{table}
%\center
%\resizebox{\textwidth}{!}{\begin{tabular}{cccccc}
%\toprule
%District & Age Group & Gender & Registered Infections  & Registration Date & Reporting Date \\
%\midrule
% \multicolumn{1}{c}{$\vdots$} & \multicolumn{1}{c}{$\vdots$}& \multicolumn{1}{c}{$\vdots$}&  \multicolumn{1}{c}{$\vdots$}& \multicolumn{1}{c}{$\vdots$}& \multicolumn{1}{c}{$\vdots$} \\
%City of Munich & 60-79 & F & 3  & June 22, 2020 & June 25, 2020 \\
%City of Munich & 60-79 & M & 5  & June 22, 2020 & June 25, 2020 \\
% \multicolumn{1}{c}{$\vdots$} & \multicolumn{1}{c}{$\vdots$}& \multicolumn{1}{c}{$\vdots$}& \multicolumn{1}{c}{$\vdots$}&  \multicolumn{1}{c}{$\vdots$}& \multicolumn{1}{c}{$\vdots$} \\
%City of Munich & 60-79 & F & 6  & June 22, 2020 & June 26, 2020 \\
%City of Munich & 60-79 & M & 5  & June 22, 2020 & June 26, 2020 \\
% \multicolumn{1}{c}{$\vdots$} & \multicolumn{1}{c}{$\vdots$}&  \multicolumn{1}{c}{$\vdots$}& \multicolumn{1}{c}{$\vdots$}& \multicolumn{1}{c}{$\vdots$}& \multicolumn{1}{c}{$\vdots$} \\
%\bottomrule
%\end{tabular}}
%\caption{Illustration of the raw data structure}
%\label{tab:data_structure}
%\end{table}

Table \ref{tab:data_structure} shows an exert of the data we are confronted with. Every morning, the database containing all registered COVID-19 infections is updated and released to the public, downloadable from the Robert-Koch-Institute's repository\footnote{\href{https://www.arcgis.com/home/item.html?id=f10774f1c63e40168479a1feb6c7ca74}{https://www.arcgis.com/home/item.html?id=f10774f1c63e40168479a1feb6c7ca74}}. 
The dataset contains, for each of the 412 districts, the cumulated number of confirmed cases of COVID-19 infections stratified by age group (00-04, 05-14, 15-34, 35-59, 60-79 or 80+) and gender, updated to that day, as well as the date of registration of each case by the local public health authorities (\textit{Gesundheitsämter}). Through the merging of daily downloads of this RKI report, we can construct  the full dataset as sketched in Table 1, where the release date is defined in the column ``Reporting Date''. This full data format is necessary to trace the reporting delay for each observation. It can sometimes indeed take several days for the data to get from the local health authorities to the nation-wide central one, and we thus define reporting delay as the number of days between registration date and reporting date. Note that since the RKI reports data every morning, all reported cases will have a delay of at least one day. The delay is especially high during weekends, a fact that we take into account in our model. Due to the delayed nature of reporting, the number of registered COVID-19 cases which refer to a specific registration date might change with the reporting date, as exemplified in Table \ref{tab:data_structure}. On September 25, the RKI has reported three registered infections of females in the age group from 60-79 living in the city of Munich, which were registered on September 22, 2020. Due to delayed reporting, this number increased to six in the report of September 26, 2020. The three newly reported cases have therefore been reported with a delay of four days. Note once again that the RKI dataset available for download only contains the information up to the current date, thus making daily downloads of the datasets necessary to determine reporting delay. %to construct the complete data as shown in Table \ref{tab:data_structure}, which allows to determine the reporting delay, we need daily downloads of the dataset, which we have available. 

%The data contain all confirmed cases of COVID-19 infections
%for each of the 412 districts in Germany, aggregated by gender and age group.
%Each data entry has a time stamp which corresponds to the
%registration date of a confirmed Covid-19 infection. Thanks to daily downloads of the data, we are also able to derive the time point of the reporting for each case. This allows us to simply calculate the reporting delay as well. This delay is present since the Robert-Koch-Institute collects the data from district-based health authorities, and it can sometimes take several days for the data to arrive to the central authority. Moreover, the RKI reports data every morning, so all reported cases will have a delay of at least one day.  The delay is especially high during weekends, a fact that we take into account in our model. 

For the sake of brevity, we here do not provide general descriptive statistics of the data, since these numbers can be easily obtained from many other sources. Among others, we refer to the RKI     webpage\footnote{\href{https://www.rki.de/EN/Home/homepage_node.html}{https://www.rki.de/EN/Home/homepage\_node.html}}, which also includes a dashboard  to visualize the data (see also CoronaMaps\footnote{\href{https://corona.stat.uni-muenchen.de/maps}{https://corona.stat.uni-muenchen.de/maps}}).

\section{Surveillance Model}  
\label{sec:surveillance}
\subsection{Framing}
\begin{figure} [h!]
	\centering
	\includegraphics[trim = 0 120 0 60, clip, width = 0.7\textwidth]{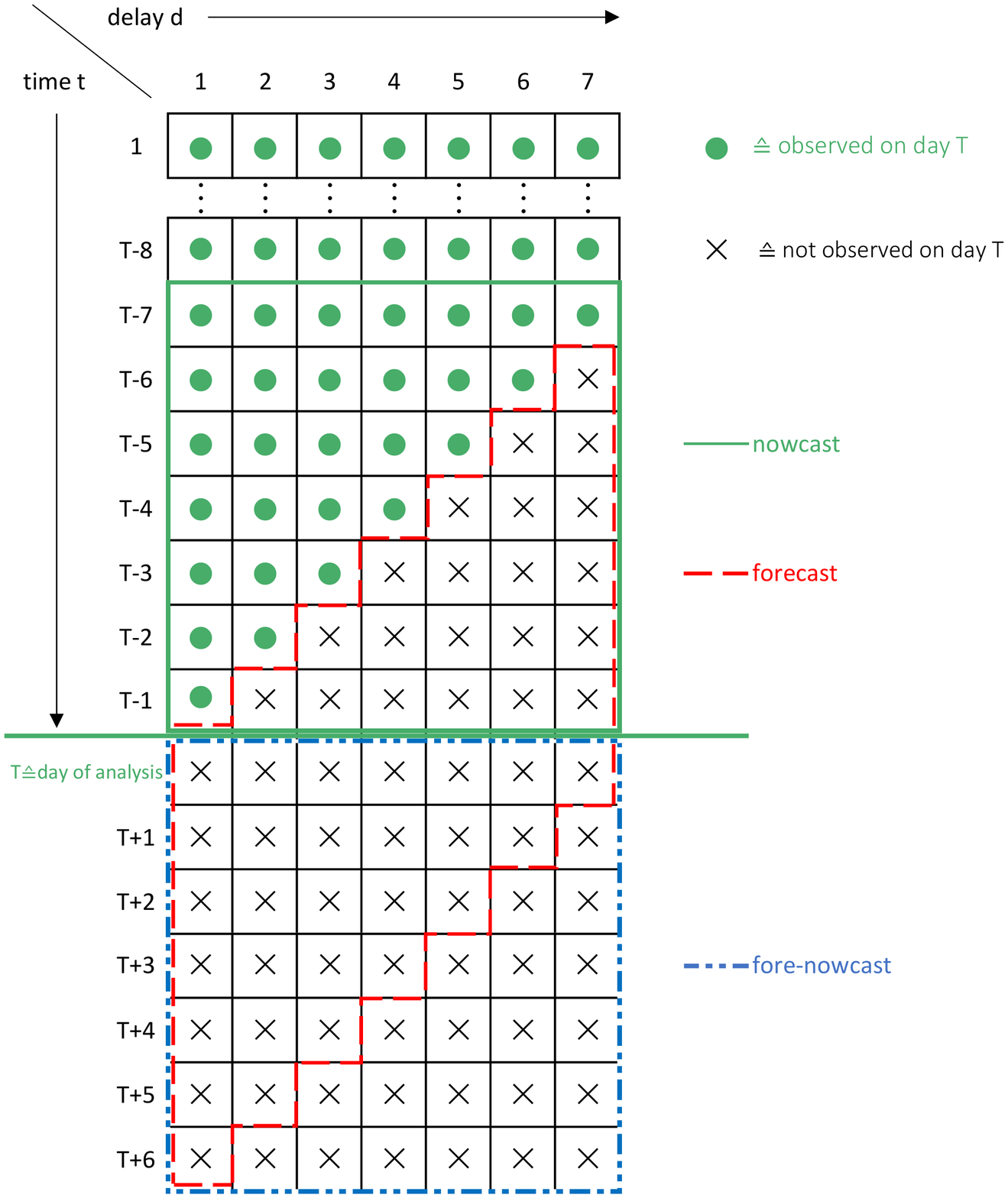}
	\caption{Sketch of the reformulated data structure showing how nowcasting, forecasting and forenowcasting are performed.}
	\label{fig:sketch}
\end{figure}

We start motivating the model by first reformulating the data structure in a way that is suitable for the analysis. Let $N_{t,d}$ denote the newly registered infections at day $t$ which are reported with delay $d$ and hence included in the database from day $t+d$.  The minimum possible delay is one day, and we assume the maximum delay to be equal to $d_{max}$ days. In our analysis we set $d_{max}=7$, which corresponds to a week. In other words, we assume delayed reporting to happen within a week.  If we define $T$ as the time point of the analysis, the data available at that moment will take the form shown in Table \ref{tab:guillotine}.
\begin{table}
\begin{center}
  \resizebox{\textwidth}{!}{\begin{tabular}{c|cccc}
   & \multicolumn{4}{c}{d}   \\
  t & 1 & 2 &  $\cdots$ &   $d_{max}$ \\
   \hline
1 & $N_{1,0}$ & $N_{1,1}$ & $\cdots$ & $N_{1,d_{max}} $\\
2 & $N_{2,0}$ & $N_{2,1}$ & $\cdots$ & $N_{2,d_{max}} $ \\
$\vdots$ &  $\vdots$ &  $\vdots$& $\vdots$ & $\vdots$  \\
$ T-d_{max}$   & $ N_{T-d_{max},0}$ &  $ N_{T-d_{max},1}$ & $\cdots$ & $N_{T-d_{max}, d_{max}}$ \\
$ T-d_{max}+1$ & $N_{T-d_{max}+1,0}$ & $N_{T-d_{max}+1,1}$ & $\cdots$ & $\mbox{NA}$ \\
$ \vdots$ &  $\vdots$ &   $\vdots$ & $\vdots$ & $\vdots$ \\
$ T-1 $& $N_{T-1,0}$ & $N_{T-1,1}$& $\mbox{NA}$  & $\mbox{NA}$ \\
$ T $& $N_{T,0}$ &$\mbox{NA}$ & $\mbox{NA}$ & $\mbox{NA}$ \\
   \end{tabular}}
   \end{center}
\caption{Reformulated data structure for a single district, age group and gender, explicitly including delay. Available data are akin to a guillotine blade.}
\label{tab:guillotine}
\end{table}
The bottom right triangle of the data is missing,  so that the structure of the available data is akin to that of a guillotine blade. 
This comparison can be helpful to understand prediction of future values, since predicting by reporting date corresponds to making the blade fall down by one or more days. In other words,  one of our goals will be to predict the diagonal edge of the blade, which corresponds to the predicted cases reported on day $T+1$. To better explain our prediction strategy, we give a sketch of this idea in Figure \ref{fig:sketch}. In the sketch, the green dots represent data that are already observed at time T (the day of analysis), while the crosses represent entries that are not yet observed and that we aim to predict with our model. This is done in three steps, which are described below. To be specific, we pursue {\sl nowcasting}, {\sl forecasting} and the combination of both, {\sl forenowcasting}.

\paragraph{Nowcasting:} Since each row of the matrix contains cases registered on a single date, to obtain the amount of cases registered on that day, regardless of the delay with which they are reported, we need to take the sum of the corresponding row. If the goal is to obtain predictions by registration date for several days, we then just sum the cases over the corresponding rows. In Figure \ref{fig:sketch} we highlight this type of prediction with a green square, which represents a weekly nowcast, that is the number of cases with registration dates over the past week. This comprises numbers that have already been observed as well as the predictions for cases from past days that have not yet been reported. 

%The blue square depicts instead a weekly \textsl{forenowcast}, that is the prediction for cases that will be registered over the next week (not all of which will be reported during said week). In this case, all entries are clearly unobserved and need to be predicted.

\paragraph{Forecasting:} If we shift the focus from predicting by registration date to reporting date, that is, if the aim is to predict reported numbers regardless of when the reported infections were actually first discovered, we cannot sum the entries of the matrix row-wise, but we need to do so diagonally. This is because the reported number on day $T$ is comprised of the sum of cases registered on day $T-1$ reported with delay 1, cases registered on day $T-2$ reported with delay 2, and so on and so forth, up until cases registered on day $T-d_{max}$ reported with delay $d_{max}$. The red parallelogram in Figure \ref{fig:sketch} thus represents the cumulated weekly forecast, that is, the predicted number of infections to be reported over the next seven days. Here all entries are unobserved, and will need to be predicted through our model, which will be uncovered in the following section.

\paragraph{Forenowcasting:} We can also combine the two aspects and predict the number of infections that will happen in the next week, regardless of their reporting date. While the previously described forecasting (i.e. predicting by reporting date) is useful to have an idea of the numbers that will be reported each day, what really gives a picture of the ongoing situation are infection numbers based on registration date. This weekly prediction corresponds to the blue square in Figure \ref{fig:sketch} and in fact is a combination of forecasting and nowcasting. We will demonstrate that the three types of predictions can be carried out with a single model, where predictions refers to extending the data row-wise (i.e. nowcasting), column-wise (i.e. forecasting) or both (i.e. forenowcasting).

%As anticipated, registered infections can also be cumulated row-wise over the delay $d$. We define those as 
%\begin{equation}
%\label{eq:Def-C}
%C_{t,d} = \sum_{j=1}^d N_{t,j}
%\end{equation}
%so that $C_{t,d_{max}}$ gives the overall number of registered infections for day $t$. This cumulated number will be used in the model.
%We do  have the same guillotine blade data structure for $C_{t,d}$ as for $N_{t,d}$, with the only difference lying in the fact that the registered infections will this time be cumulated row-wise. 

%As explained in the Data section, 
%new infections are detected through tests conducted by the local health authorities. These autorities then report the numbers to the RKI with an inevitable delay, leading to the data structure described above, with the matrix being available at the district level. Moreover, 

\subsection{Statistical Model}

As already stated in Section \ref{sec:data}, the cumulative numbers of registered COVID-19 infections are, other than by registration date, also stratified by district, age group and gender.  To accommodate this additional information, we extend the notation from above and define with $N_{t,d,r,g}$ the number of newly registered infections on day $t$ in region/district $r$ and gender and age group $g$, reported by the RKI on day $t+d$ (thus with delay $d$). Row-wise cumulated numbers are defined through 

\begin{equation}
\label{eq:Def-C}
C_{t,d,r,g} = \sum_{j=1}^d N_{t,j,r,g}
\end{equation}
which represents the group- and district-specific cumulated number of cases with registration date $t$ and delay up to $d$. We define with $\boldsymbol{z}_r$ the geo-coordinates of district/region $r$ and generally denote covariates with $\boldsymbol{x}$, where varying subscripts indicate dependence on either gender- and age group $g$, region $r$, time point $t$ or delay $d$. 

We assume the counts $N_{t,d,r,g}$  to follow a negative binomial distribution with mean $\mu_{t,d,r,g}$ and variance $\mu_{t,d,r,g} + \theta \mu_{t,d,r,g}^2$, where $\theta>0$ and the limit $\theta \rightarrow 0$ leads to a Poisson distribution. More specifically, we set

%Note that $d$ takes values $d=1,  \ldots, d_{max}$. 

%For the mean structure we condition on all available data from the previous day, which incorporates  the epidemic component of the model. To be specific, we set
%\begin{align}
\label{eq:model}
\begin{equation}
\begin{split}
 \mu_{t,d,r,g} = \exp\{ s_1(t) + s_2(\boldsymbol{z}_r) + {\gamma}_d + \boldsymbol{x}_{t,d} \boldsymbol{\alpha} +  
\boldsymbol{x} _{g} \boldsymbol{\beta} + {\bf x}_{t} {\bf u}_r  + \\ +  \phi \log(1 + C_{t-1,d,r,g}) + \delta \log(1 + C_{t,d-1,r,g}) + \mbox{offset}_{r,g}\} 
\end{split}
\end{equation}
%\end{align}
%+ \log(C_{t,d-1,r,g}) \delta
Here, $s_1(t)$ is a global smooth time trend and $s_2(\boldsymbol{z}_r)$ is a smooth spatial effect over the districts of Germany. 
The parameters $\boldsymbol{\gamma}_d = (\gamma_1, \ldots , \gamma_{d_{max}})$ capture the delay effect for each delay $d$, while the parameters contained in $\boldsymbol{\alpha}$ capture effects related to time and delay, which in our case will be weekday effects. Gender and age effects are included in ${\boldsymbol{\beta}}$, and ${\bf u}_r$ are unstructured regional effects which will be subsequently specified in more detail. Coefficient $\phi$ captures the time-related autoregressive (AR) component of the process, indicating the effect of cases from the same district and gender- and age group which were reported on the previous day. Coefficient $\delta$ expresses the effect of infections registered on the same day which were reported with delay up to $d-1$, or in other words a delay-related autoregressive component. Finally, the offset is set to the logarithm of the regional population size in the different gender- and age-groups, enabling us to model the infection rate. The offset defined this way also allows to incorporate the size of the susceptible population in each region, showing that this type of modeling is practicable at  different stages of the pandemic. In this case, the population size would need to be replaced by the number of susceptible in region $r$, incorporating the SIR (susceptible-infected-removed) model or other similar ones (see e.g. \citealp{allen1993}). This is not particularly relevant at the time point chosen for the analysis, since
the number of susceptible corresponds more or less to the population size due to the small (and unknown) size of the immune populations in each district.

The previously mentioned spatial effect is comprised of two components: An overall smooth effect $s_2(\boldsymbol{z}_r)$ mirroring the fact that different parts of Germany are differently affected, and a region-specific component accounting for  infection rates that are particularly high or low in single districts with respect to the neighbouring situation. To be more specific, $s_2()$ is a smooth spatial function of the geo-coordinates $\boldsymbol{z}_r$ for region $r$, while the $\boldsymbol{u}_r$ are unstructured region-specific effects, interacting with the time dependent covariates
$\boldsymbol{x}_{t}$. 
We put a normal prior on $\boldsymbol{u}_r$, i.e. we model $\boldsymbol{u}_r = (u_{r0}, u_{r1})^\top$ as random effects, where
${u}_{r0}$ is a general random intercept capturing the long-term level (from $t = 1,\dots,T$) of the epidemiological situation in the different districts, while ${u}_{r1}$ is a second random intercept estimated exclusively over the last $k$ days, expressing the short-term dynamics (within $k$ days prior to $t = T$) of infections. In our analysis we set $k=7$. For $\boldsymbol{u}_r$ we assume the structure
\begin{equation}
\boldsymbol{u}_r \overset{iid}{\sim} N(\boldsymbol{0}, \boldsymbol{\Sigma}_u) %\mbox{ i.i.d}
\label{eq:u_r}
\end{equation}
for $r = 1,\dots,412$, with the posterior variance matrix $\boldsymbol{\Sigma}_u$ being estimated from the data. 
The predicted values $\widehat{\boldsymbol{u}}_r$ (i.e. the posterior mode) measure how much and in which direction the infection rate of each district deviates from the global spatial structure, controlling for covariates and age- and gender-specific population sizes.
%\textcolor{red}{GK: Isn't this a rather relevant result, which I would include.
%Moreover, we could include autoregressive components in the model, for
%instance we could include $log(C_{t-1,d,r,g})$ or even some variant of that, e.g.
%$C_{t-1,d,r,\cdot}$, where we sum over all age groups. Therefore, I propose to erase the subsequent paragraph}.
%Note that the model does not include $C_{t,d-1,r,g}$, the group- and region-specific cumulated number of infections with delay up to $d-1$. This is because we found that, given all information that is already included in the model, it did not substantially increase the accuracy in the nowcasting and forecasting of infections $N_{t,d,r,g}$. 

\subsection{Model Selection and Performance}
\label{sec:evaluation}

Model (\ref{eq:model}) includes several components. In this section we aim at assessing whether the inclusion of some of those components is beneficial in terms of predictive performance, and to generally evaluate the overall performance of the final model. We are specifically interested in seeing how the unstructured random effects ${\bf x}_{t} {\bf u}_r$ and the autoregressive components $\phi \log(1 + C_{t-1,d,r,g})$ and $\delta \log(1 + C_{t,d-1,r,g})$ impact predictive accuracy. To do so, we consider the realized absolute prediction error with regards to nowcasts, forecasts and forenowcasts, cumulated for each district over a period of seven days using different model specifications, to compare performance over time through a weekly rolling window approach. The specifics of how predictions are performed will be described in detail in Section \ref{sec:predictions}.

Starting with nowcasting, let therefore  $Y_{T,r}^{(n)} $ denote the cumulated number of registered infections in district $r$ over $k = 7 $ days prior to the day of analysis at time $T$, that is 
\begin{equation}
	\nonumber
	%Y^{(n)}_{T,r} = \sum_{t=1}^{d_{max}} \sum_g C_{T-t,d_{max},r,g}
	Y^{(n)}_{T,r} = \sum_{t=1}^{k}\sum_{g} C_{T-t,d_{max},r,g}
\end{equation}
with ${C_{T,d_{max},r,g}}$ defined as in (\ref{eq:Def-C}). This corresponds to the sum of all numbers in the green square in Figure \ref{fig:sketch}. Accordingly, we define with $\widehat{Y}^{(n)}_{T,r}$ the corresponding prediction based on the fitted model as described above. For forecasting, we modify the definition and  look at the cumulated number of cases
\begin{equation}
	\nonumber
	Y_{T,r}^{(f)} = \sum_{t = 1}^{k} \sum_{d=1}^{d_{max}} \sum_g N_{T+t-d,d,r,g}
\end{equation}
which corresponds to the red parallelogram in Figure \ref{fig:sketch}. Again, the corresponding predicted value is notated as 
$\widehat{Y}_{T,r}^{(f)}$. Finally, for forenowcasting we concentrate on  the cumulated numbers in the blue square, and set 
\begin{equation}
	Y^{(fn)}_{T,t} = \sum_{t=1}^{k}  \sum_g C_{T+t-1,d_{max}, r,g}
	\nonumber
\end{equation}
with matching prediction $\widehat{Y}^{(fn)}_{T,t} $ based on the fitted model. With the notation just given, we can define the relative district specific prediction error (standardized per $100,000$ inhabitants) simply as 
\begin{equation}
	\nonumber
	\text{RPE}_{T,r}^{(\cdot)} = 100,000 \frac{ Y_{T,r}^{(\cdot)} -  \widehat{Y}_{T,r}^{(\cdot)}}{\text{pop}_{r}}
\end{equation}
where $pop_r$ is the population size in district $r$, and the dot refers to nowcasting, forecasting or
forenowcasting, respectively. It should be clear that, setting $k = d_{max} = 7$, the numbers defined above are only observable on day $T+7$ for nowcasting
and forecasting and on day $T+14$ for forenowcasting.

To obtain a measure of the overall predictive performance of the model for a certain fitting date $T$, we take the mean of $\text{RPE}_{T,r}$ in absolute value over all districts, which we call Mean Absolute Relative Prediction Error (MARPE):
\begin{equation}
	\nonumber
	\text{MARPE}_{T}^{(\cdot)} = \frac{1}{412}\sum_{r = 1}^{412} {|\text{RPE}_{T,r}^{(\cdot)}|}
\end{equation}
To have an idea of the average bias of predictions over time, we also plot the Mean Relative Prediction Error (MRPE), which takes the mean of relative errors without considering them in absolute value:

\begin{equation}
	\nonumber
	\text{MRPE}_{T}^{(\cdot)} = \frac{1}{412}\sum_{r = 1}^{412} {\text{RPE}_{T,r}^{(\cdot)}}
\end{equation}
This last measure will be positive if the model tends to underpredict on average over the districts, and negative otherwise.

To evaluate the predictive accuracy of different model specifications, we compute $\text{MARPE}_{T}^{(\cdot)}$ and $\text{MRPE}_{T}^{(\cdot)}$ over time by fitting the model weekly for each of the considered specifications in a rolling window approach. In particular, we consider te following model variations:

\begin{itemize}
	\setlength\itemsep{0em}
	\item Full model as in (\ref{eq:model});
		\item Model without the time-related autoregressive component, $C_{t-1,d,r,g}$.
			\item Model without the delay-related autoregressive component, $C_{t,d-1,r,g}$;
				\item Model without the autoregressive components, $C_{t-1,d,r,g}$ and $C_{t,d-1,r,g}$;
	\item Model without the short-term district-specific random intercept, $u_{2,r}$;
		\item Model without the unstructured district-specific random effects ${\bf u}_r$;
	\item Model without the short-term district-specific random intercept, $u_{2,r}$ and the autoregressive components, $C_{t-1,d,r,g}$ and $C_{t,d-1,r,g}$;
		\item Model without the unstructured district-specific random effects ${\bf u}_r$ and the autoregressive components, $C_{t-1,d,r,g}$ and $C_{t,d-1,r,g}$;

\end{itemize}

\begin{figure}[h!]
	\centering

\includegraphics[width =0.31\textwidth]{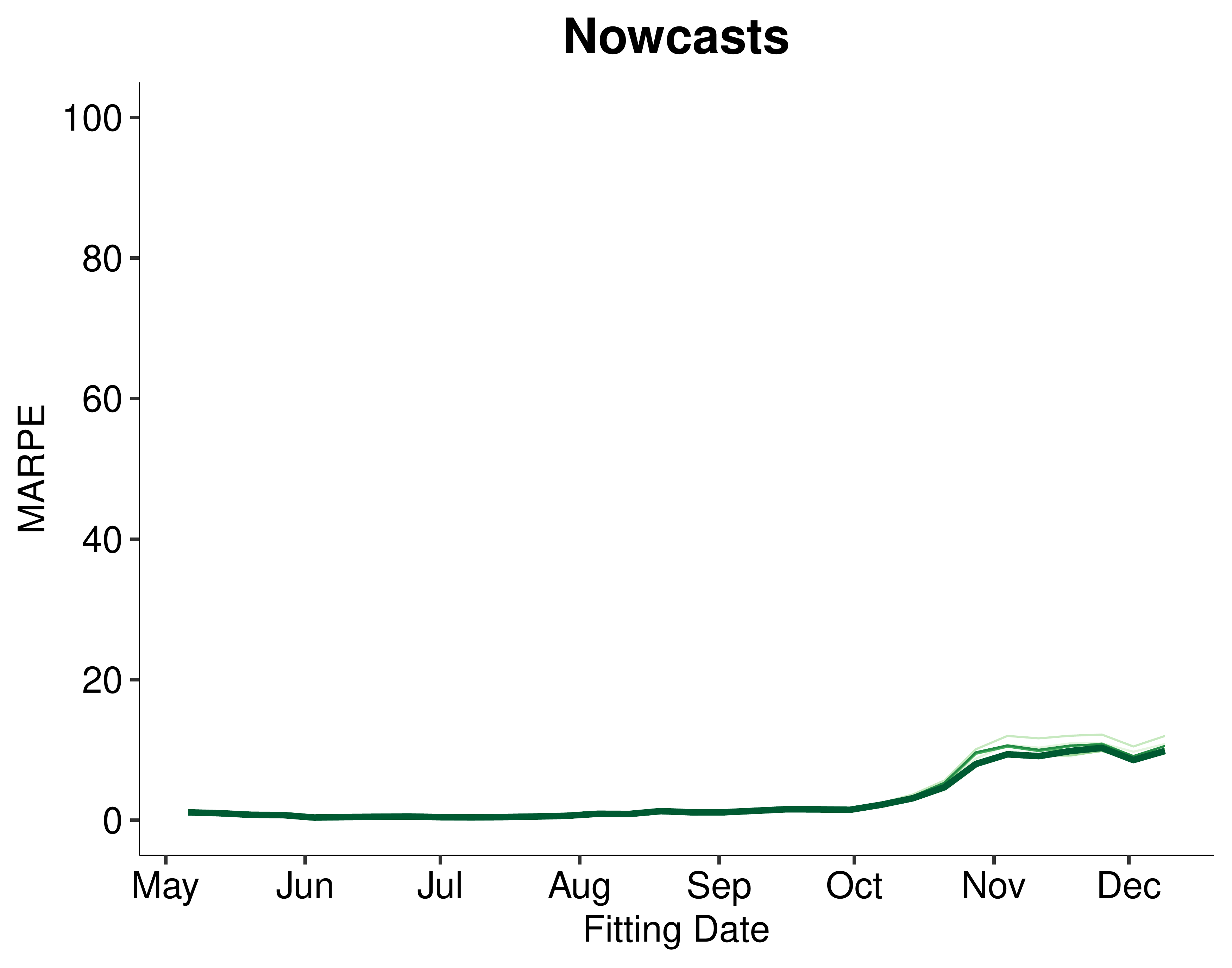}
\includegraphics[width =0.31\textwidth]{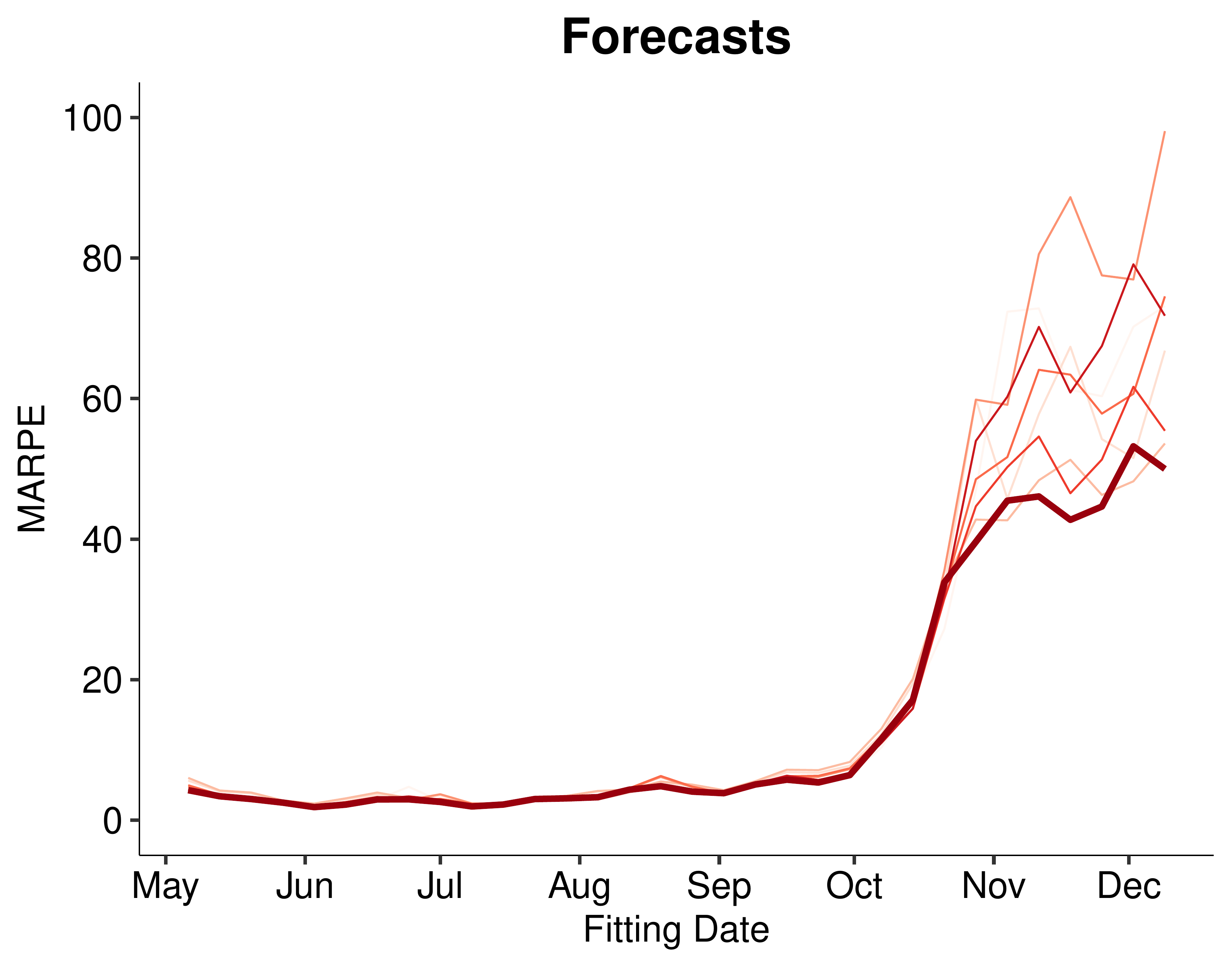}
\includegraphics[width =0.31\textwidth]{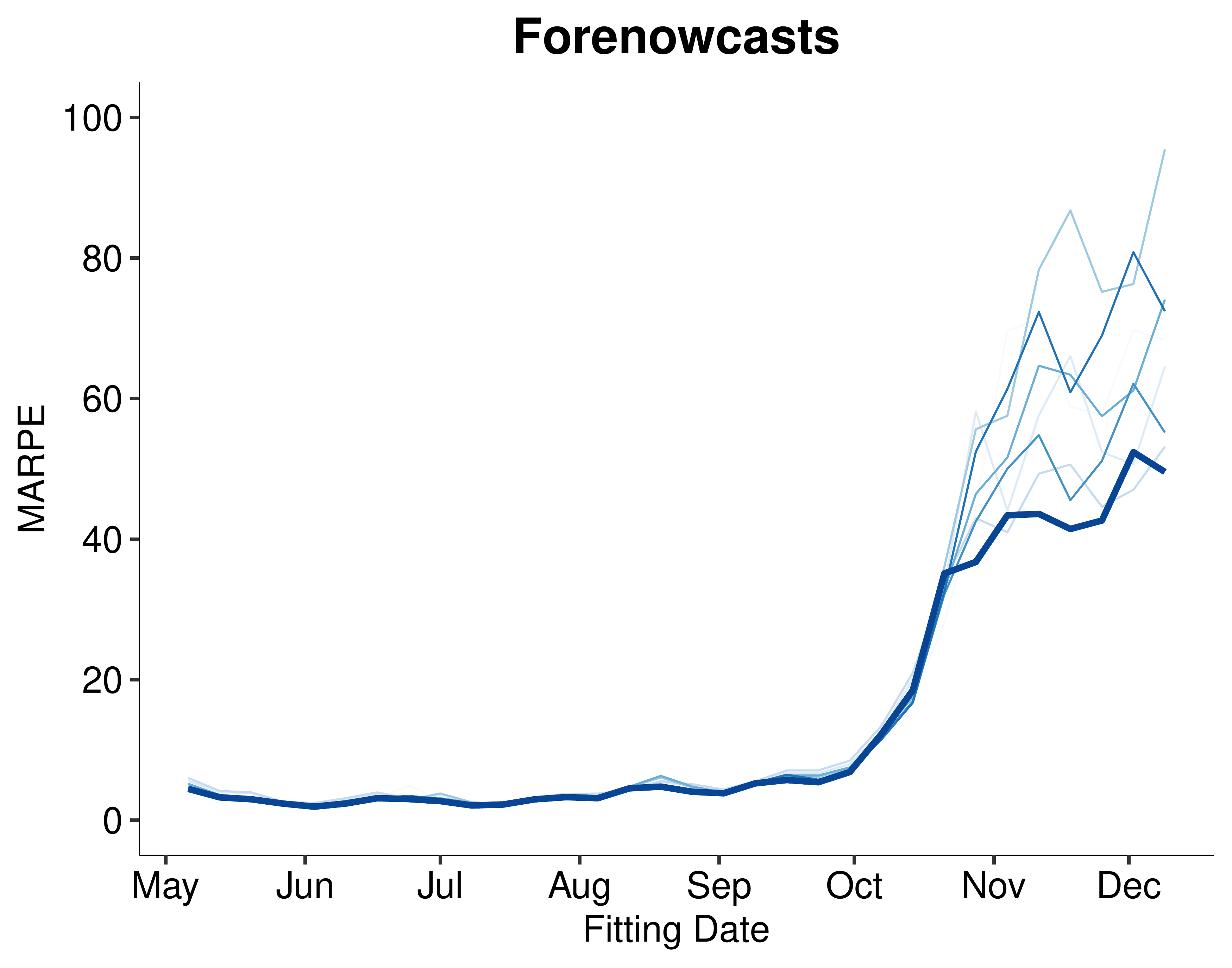}

\includegraphics[width =0.31\textwidth]{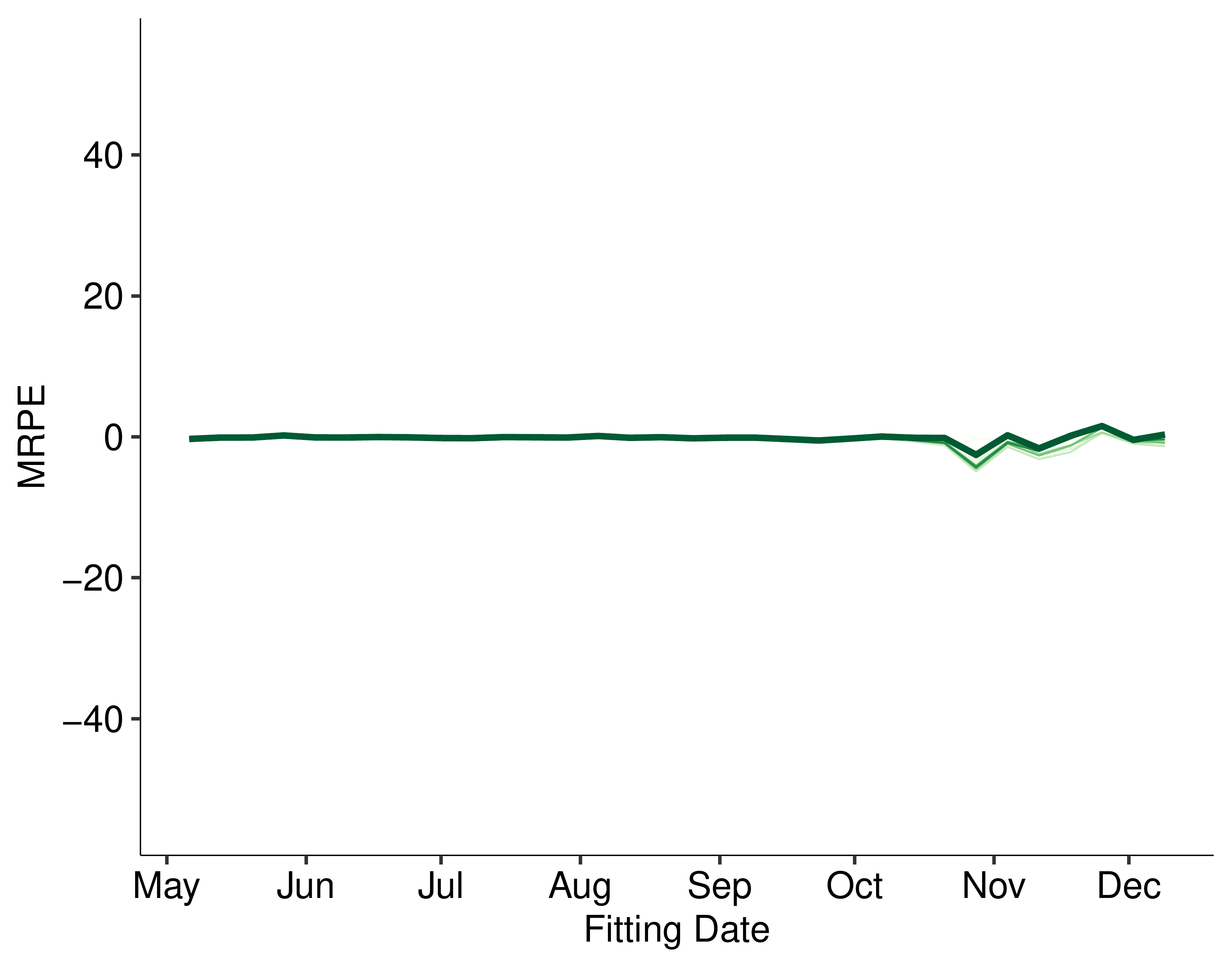}
\includegraphics[width =0.31\textwidth]{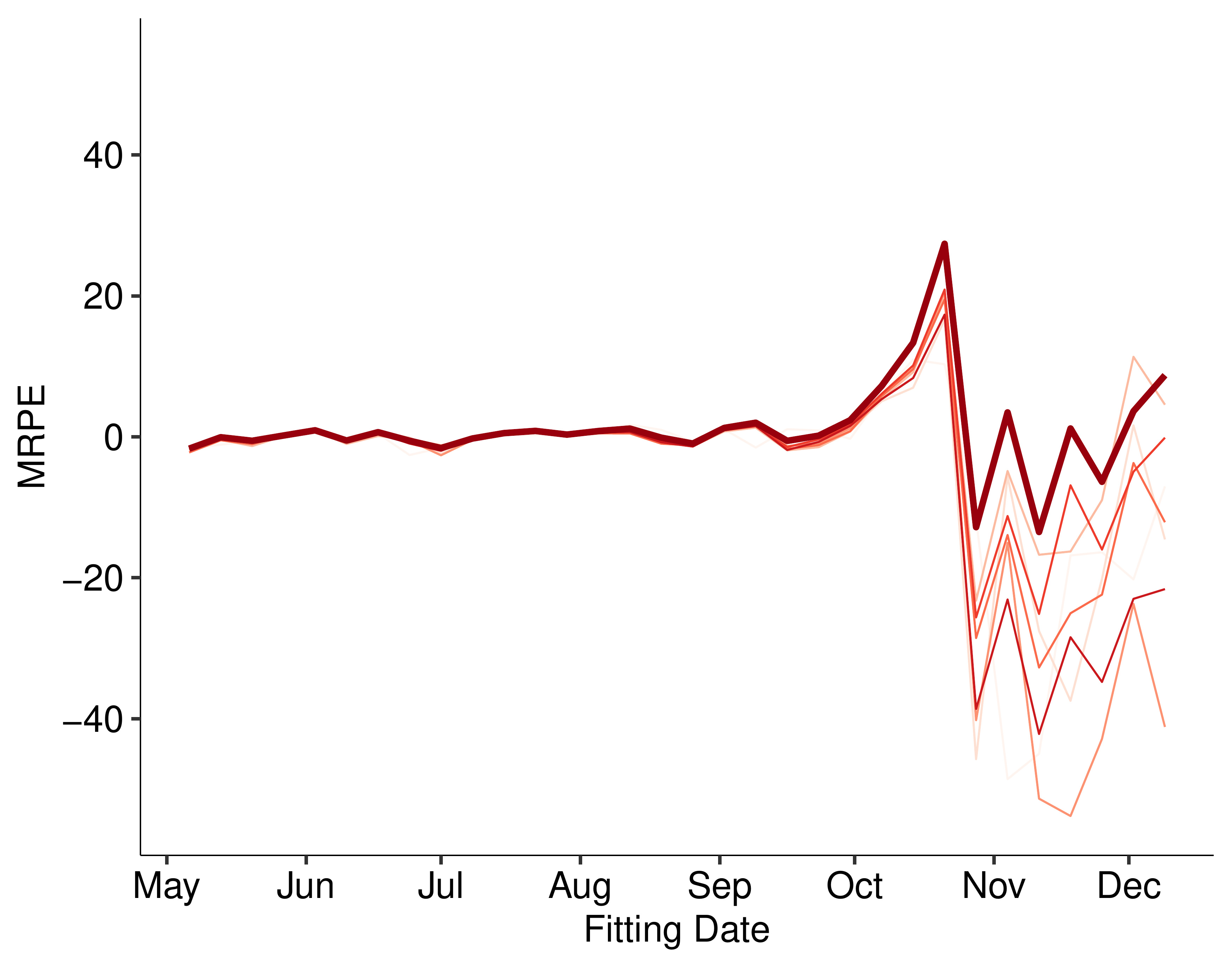}
\includegraphics[width =0.31\textwidth]{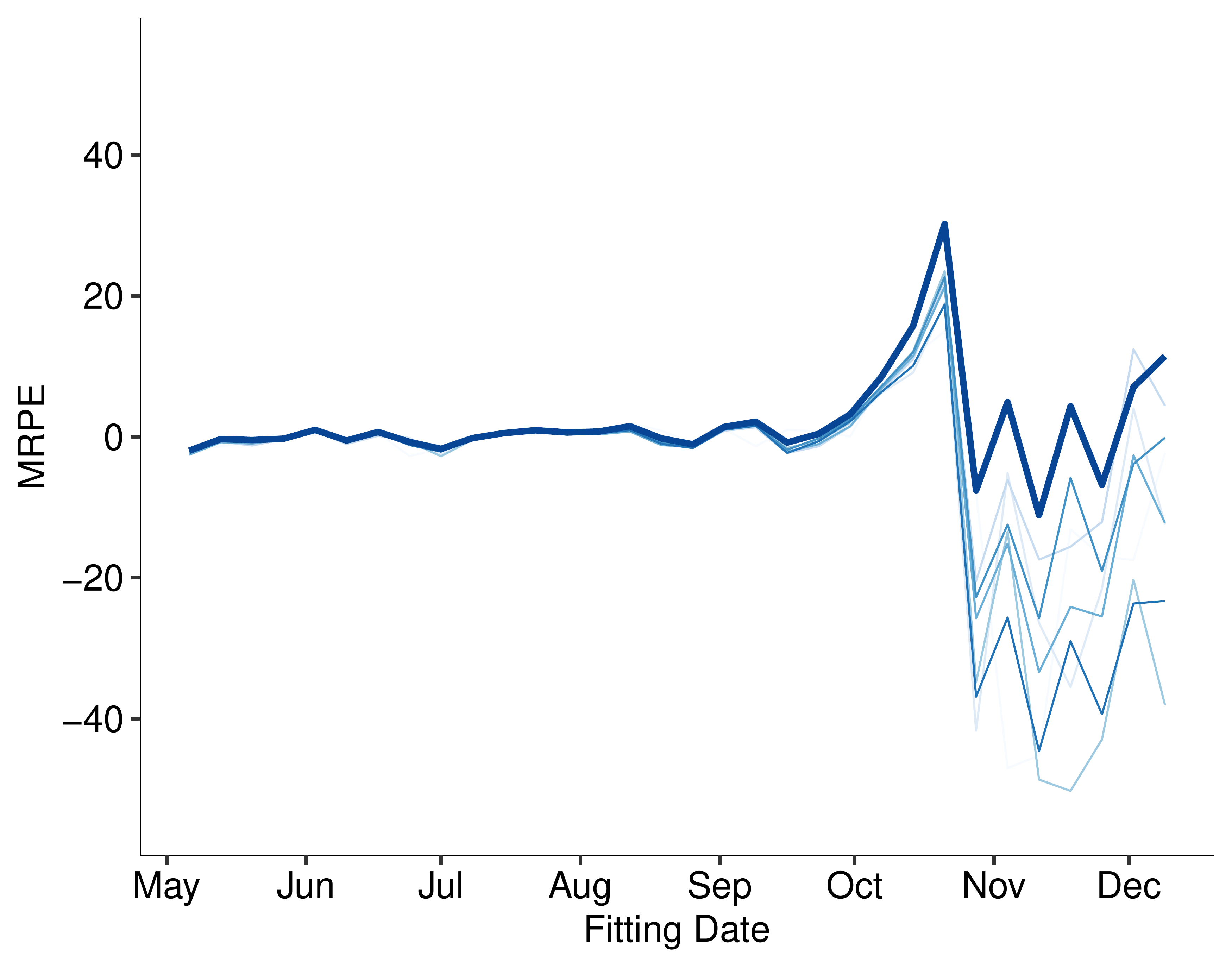}
	\caption{Mean Absolute Relative Prediction Error ($\text{MARPE}_{T,r}^{(\cdot)}$, top panel) and Mean Relative Prediction Error ($\text{MRPE}_{T,r}^{(\cdot)}$, bottom panel) for all districts in Germany calculated over time for different model specifications, respectively for nowcasts (green), forecasts (red) and forenowcasts (blue). The thicker line indicates the selected model, which corresponds to the full model with the exclusion of the time-related AR component, $C_{t-1,d,r,g}$
		.}
	\label{fig:errors}
	
\end{figure}

Figure \ref{fig:errors} plots the MARPE and the MRPE by model fitting date for nowcasts, forecasts and forenowcasts, respectively. The plots already reveal several aspects of the goodness of fit of our model.  Looking at the MARPE (top panel), we at first notice how the errors for nowcasts are, as expected, much smaller than for forecasts and forenowcasts. Secondly, we can see how prediction errors are remarkably small for the first five months of model fitting. Those  months coincide with the late spring and summer months, during which the infection numbers were relatively stable and under control in Germany. Our model was thus able to capture most of the variability in the process, resulting in precise predictions not only for nowcasts, but also for forecasts and forenowcasts. Finally, we notice how there is a large increase in MARPE for all fitted models starting from October, which coincides with the beginning of the second wave of COVID-19 in Germany. This is due to the fact that in that period infection dynamics changed and the numbers got much larger, thus also leading to an increase in prediction errors. The model variant that performed the best during this later period is the full model with the exclusion of the time-related autoregressive component $C_{t-1,d,r,g}$, highlighted with a thicker line. The plots for the MRPE (bottom panel) confirm this fact and help explaining the reasons behind it. For both forecasts and forenowcasts, we see how at the very beginning of the second wave all models tend to underpredict, while they overpredict from November onward. The chosen model without the AR component is actually the most biased one towards the downside in October (even though it is not performing worse than the others in terms of MARPE), while it then becomes by far the least biased towards the upside in later months. This is beacuse infection numbers grew very fast in October, and models including the autoregressive component were better able to capture the quick increase. In contrast though, after new infections somewhat stabilized, the models including the autoregressive component were still projecting the increase of past months on new ones, causing large overestimation. The chosen model is instead more conservative in its predictions, resulting in better overall predictive performance.

%From the plots we immediately see how there is a  fundamental change in the process starting from the end of September, which marked the beginning of the second wave in Germany. 

\section{Applied Surveillance}
\label{sec:predictions}
Given that what we propose is a monitoring tool, the results change over time. We here give an exemplary snapshot of the estimates and how predictions can be obtained using Tuesday, September 15, 2020 as date of the analysis. %providing predictions as well as infection numbers observed \textsl{a posteriori}, enabling comparison and model evaluation. 
As an additional remark, note that our analysis is completely reproducible for different dates as well, with code and data openly available and downloadable from our GitHub repository\footnote{\href{https://github.com/gdenicola/Now-and-Forecasting-COVID-19-Infections}{https://github.com/gdenicola/Now-and-Forecasting-COVID-19-Infections}}.

\subsection{Model-Based Monitoring}

\begin{figure}[]
	\center
	\begin{subfigure}[b]{0.46\textwidth}
		\caption*{}
		\includegraphics[width =\textwidth]{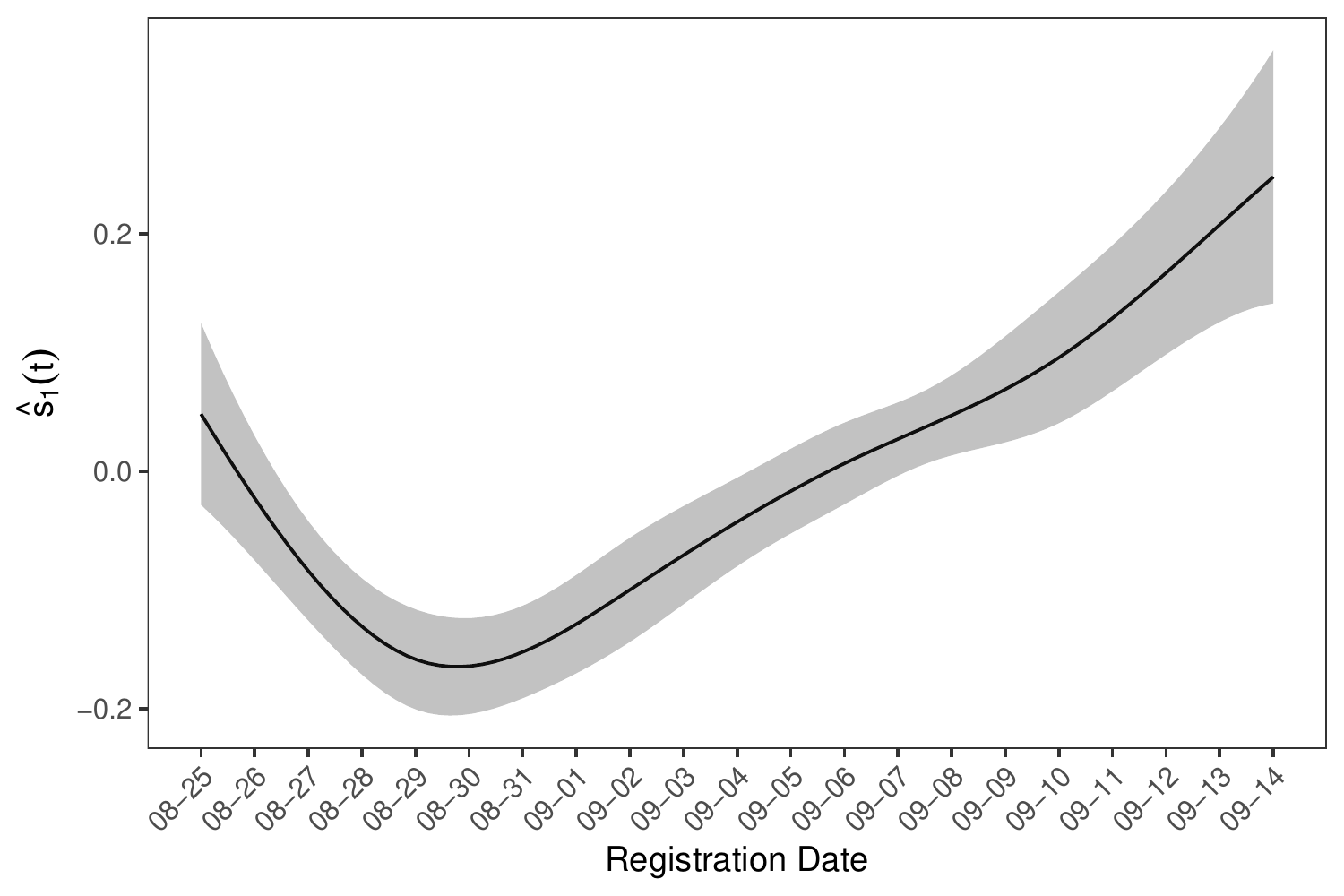}
		\label{fig: smooth time effect}
	\end{subfigure} 
	\begin{subfigure}[b]{0.44\textwidth}
		\caption*{}
		
		\includegraphics[width = \textwidth]{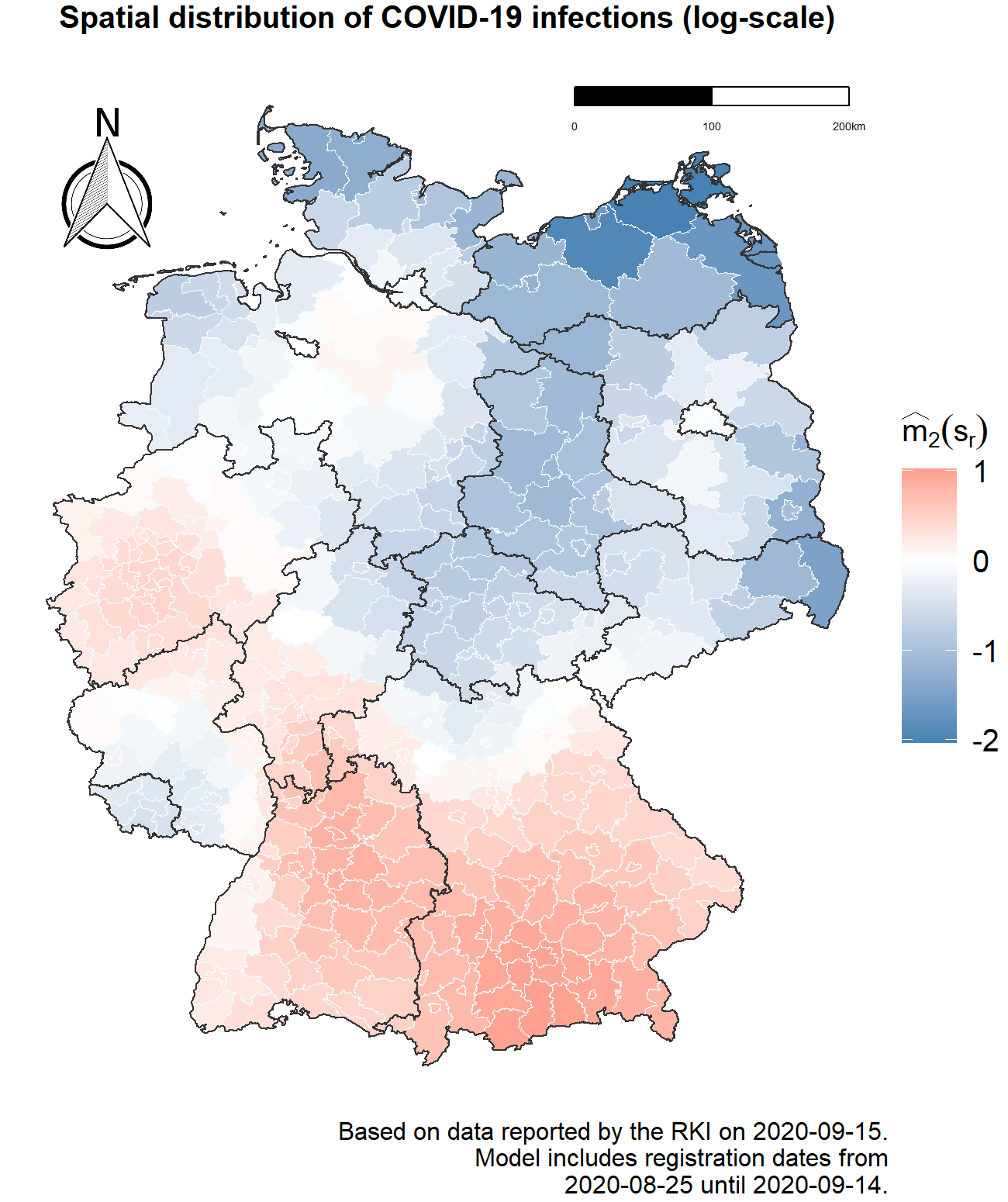}
		\label{fig: smooth spatial effect}
	\end{subfigure}
	\caption{Estimated smooth effects $s_1(t)$ and $s_2(\boldsymbol{z}_r)$, respectively the fitted smooth effect of time and the fitted smooth spatial effect for the prevalence of COVID-19 infections in Germany (measured on the log scale). Both effects are estimated over the 21 days prior to September 15, 2020.}
	\label{fig:smooths}
	
	\medskip

\includegraphics[width = 0.45\textwidth]{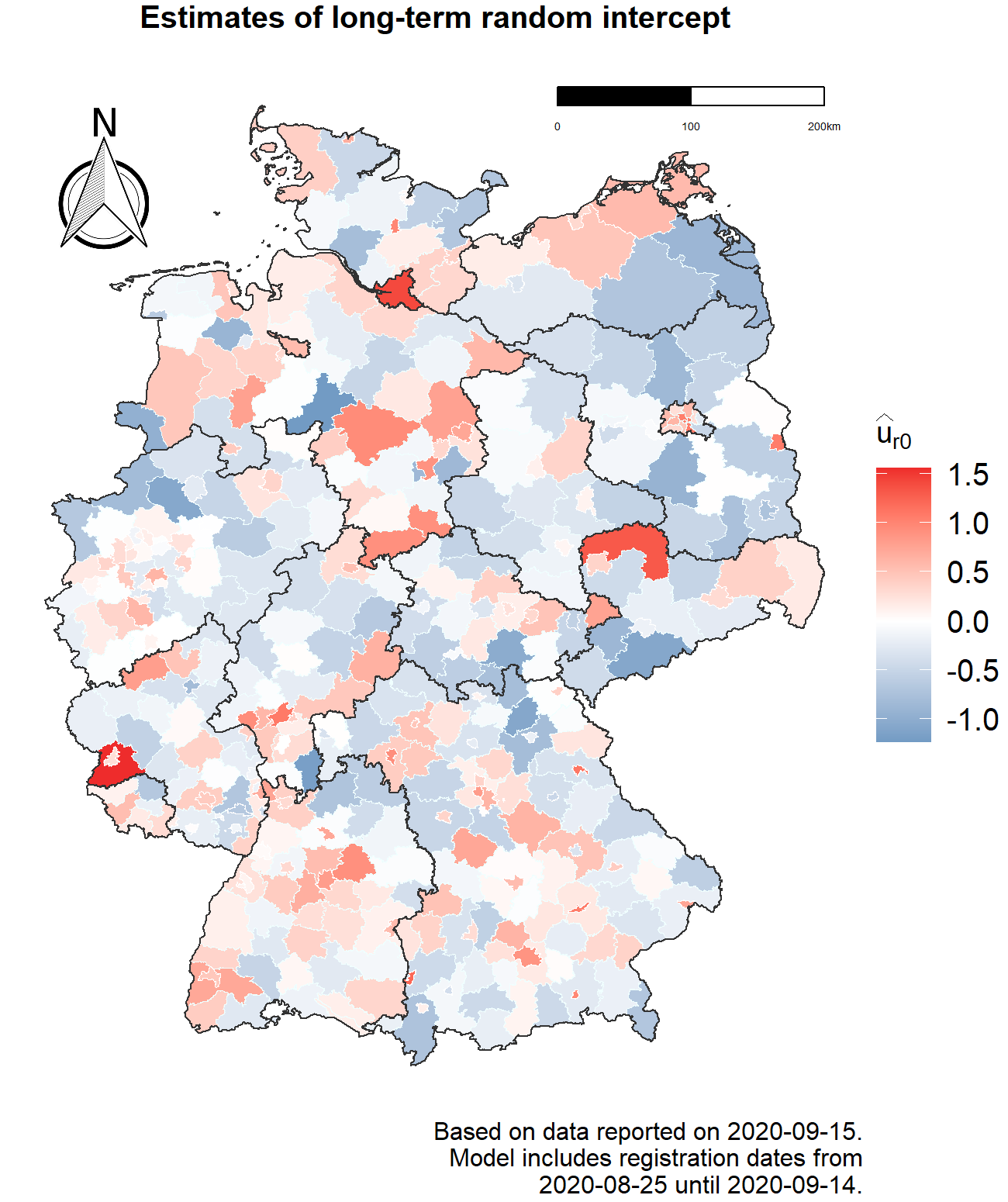} 
\includegraphics[width = 0.45\textwidth]{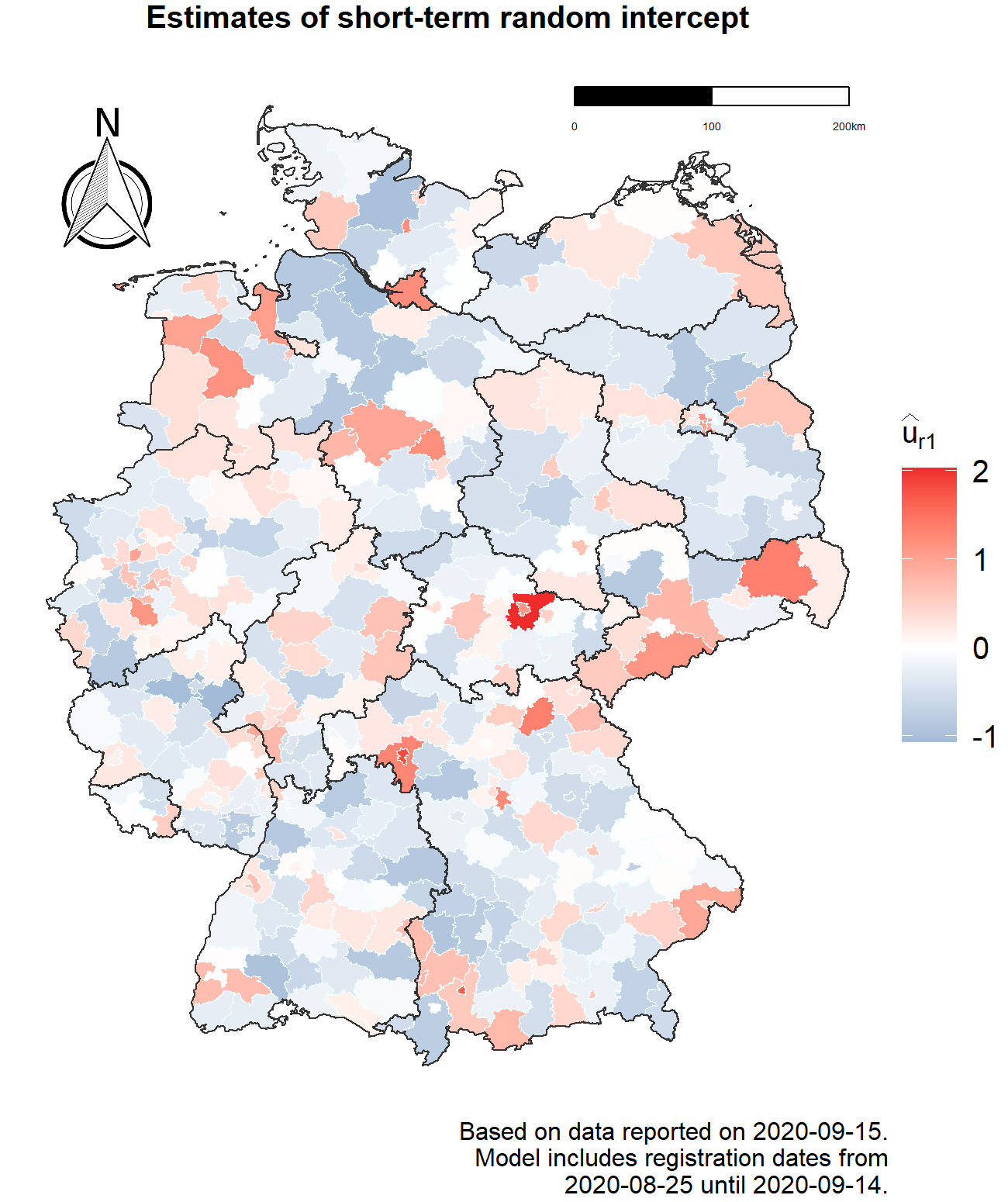} 
\caption{Region specific level (left) and dynamics (right) of  COVID-19 infections, controlling for the smooth spatial effect on the right hand side of Figure \ref{fig:smooths}.}
\label{fig:randomintercepts}

\end{figure}

In addition to proper predictions (nowcasts, forecasts and forenowcasts), which will be shown in the next section, our model also fits smooth components over time and space, which are visualized in Figure \ref{fig:smooths}. The left hand side shows the estimated infection rate over time for the three weeks prior to \doa. We notice how the rate of registered infections has been dropping until the end of August, while in the following weeks numbers started rising again, leading to an inversion and a steady increase in the smooth spline.
%The spatial components also play a fundamental role in pandemic surveillance.  
The map on the right hand side depicts the smooth spatial effect estimated as a function of longitude and latitude, on the log scale. From the plot we can see how, at the time of the analysis, the regions of Bavaria and Baden-Württemberg in the south of Germany were generally the most affected. We also observe that the west was also, on average, more affected than the east.  %To give an idea of the heterogeneity that is present across Germany, we can observe that, controlling for all other factors, the infection rate is approximately 20 times higher in the most affected regions (Eastern Bavaria and Western North-Rein-Westphalia) than in the least affected regions (the north-easternmost districts of Mecklenburg-Western Pomerania). 

The two maps in Figure \ref{fig:randomintercepts} show further spatial components of the model, namely the district-specific random intercepts. Those reflect the situation in single districts controlling for the previously shown smooth spatial effect, that is, in comparison to the average of the neighboring areas. More specifically, the map on the left displays the overall district-specific long-range random intercept, depicting the relative infection situations in the 21 days prior to the day of analysis, while the map on the right hand side shows the additional short-term random intercept which enters the linear predictor only over the last 7 days, giving an idea of the more recent infection dynamics. We can thus see that, for example, the district of Weimarer Land in the region of Thuringia has had the most rapidly evolving number of cases in the 7 days prior to the day of analysis controlling for the situation in its surroundings, reflecting the outbreak that happened in the region during the analyzed period. %\citep{Middletonm2716}
This second map can already be regarded as a first way of monitoring infection dynamics at a local level, preliminary to looking at the predicted numbers: If a district has a very high short term random effect, it probably means that something is going on there that deserves further consideration. 

%In the following week, a very large number of cases were detected in an outbreak in slaughterhouses in the district of Gütersloh, leading to an erratic increase in newly registered COVID-19 infections. Since this outbreak  hardly spread over to neighbouring districts, the rate dropped down again towards the end of June.

 %We can see how the rate steadily goes down over the considered time period, showing how the restrictive measures have been able to contain the contagion thus far.

%\begin{figure}[t]
%	\includegraphics[width = 0.51\textwidth]{../../Plots/cases/TimeEffect/2020-06-29.pdf} 
%	\includegraphics[width = 0.51\textwidth]{../../Plots/cases/SpatialEffect/2020-06-29.png} 
%	\caption{Region specific level (left) and dynamics (right) of  COVID-19 infections, controlling for the spatial effect in Figure \ref{fig:spatialeffect}.}
%	\label{fig:randomintercepts}
%\end{figure}

%make this 0.4 for single page with figure 4

%\begin{figure}[t]
%	\includegraphics[width = \textwidth]{../../Plots/cases/TimeEffect/2020-06-29.pdf} 
%	\caption{Fitted smooth effect of time $s_1(t)$ over the 21 days prior to the analysis.}
%	\label{fig:time trend}
%\end{figure}

%\begin{figure}[t]
%\center
%	\includegraphics[width = 0.7\textwidth]{../../Plots/cases/SpatialEffect/2020-06-29.png} 
%	\caption{Fitted smooth spatial effect for the prevalence of COVID-19 infections in Germany (measured on the log scale).}
%	\label{fig:spatialeffect}
%\end{figure}

%make this 0.42 for single page with figure 3
%\begin{figure}[]
%	\includegraphics[width = 0.51\textwidth]{../../Plots/cases/RandomInterceptLong/2020-09-15.png} 
%	\includegraphics[width = 0.51\textwidth]{../../Plots/cases/RandomInterceptShort/2020-09-15.png} 
%	\caption{Region specific level (left) and dynamics (right) of  COVID-19 infections, controlling for the smooth spatial effect on the right hand side of Figure \ref{fig:smooths}.}
%	\label{fig:randomintercepts}
%\end{figure}

\subsection{Predictions}

As previously explained, our model can be used to directly nowcast (correct reports from previous days for delay), forecast (predict the number of cases reported in the next days) and forenowcast (predict the number of infections that will be registered for the next days). The obtained predictions can be used to get a picture of how the pandemic is going to unfold in the short term. In the following, we explain how we obtain those predictions from our model. 

\paragraph{Nowcasting:}
%We can use our model to nowcast the missing observations from past days which have not yet been observed due to the reporting delay. Other than  to obtain a more complete picture of the current epidemiological situation, this also allows for an up-to-date computation of dynamics measures such as the 7-days incidence. 
In our case, nowcasting is equivalent to filling all NA (missing) entries of the matrix in Table \ref{tab:guillotine}, turning the trapezoid shape of the data into a full rectangle. This is also equivalent to completing the green square in Figure \ref{fig:sketch}. Given that we model delay $d$ as a stand-alone variable in our generalized additive model, we are able to simply predict the missing cells directly by setting the delay $d$ to the necessary value  in the data vector used for predictions alongside all other covariates. We can thus nowcast infections for each delay, day, district, gender and age group. 
The predictions are, depending on the model variant used, also dependent on the autoregressive terms $C_{t-1,d,r,g}$ and $C_{t,d-1,r,g}$. These are, except when predicting the first diagonal of the red parallelogram in Figure \ref{fig:sketch}, not yet known at the day of analysis.  We therefore perform the prediction of the black crosses in Figure \ref{fig:sketch} iteratively, by utilizing the predictions of the previous diagonal as the autoregressive components. 
Based on the model, we can also take uncertainty into account by simulating data from a negative binomial distribution with the corresponding mean and variance structure. More precisely, we apply the same strategy as above, but instead of using the mean value we now plug counts simulated from the model into the autoregressive components, and repeat this procedure $n = 1000$ times. This parametric bootstrap approach easily allows us to compute lower and upper bounds of the prediction intervals.

\paragraph{Forecasting:} The model also allows to directly predict cases for future dates. With $T$ denoting the time point of data analysis, we can obtain predictions for the number of reported cases on days $T, T+1, \ldots T+k-1$. Let us start with the predictions for the reported cases on the day of analysis, i.e.\ time $T$. Referring once again to the guillotine blade structure in Table \ref{tab:guillotine}, we proceed as follows: For $d=1$, i.e.\ at the leftmost point of the blade, we take the fitted mean values as prediction, while keeping the smooth function of time constant, that is, setting $s(t+1) \equiv s(t)$ for the sake of stability. For the remaining $d_{max}-1$ elements of the blade edge we take the mean value by setting $d=d+1$. To get predictions for the numbers of infections reported on days $T+1, \ldots T+k-1$ we can then proceed in an analogous way, using the values just predicted to update the autoregressive components ($C_{t-1,d,r,g}$ and $C_{t,d-1,r,g}$).
Figure \ref{fig:sketch} visualizes the strategy, with cumulated predictions for the number of cases reported on days $T,T+1,\ldots T+6$ being represented by the red parallelogram. Similarly as we did for the nowcasting, we can take uncertainty into account through simulations, sampling from a negative-binomial model with the estimated group-specific mean and variance structure.

\paragraph{Forenowcasting:} While the previously described forecasts (i.e. predictions by reporting date) are useful to have an idea of the numbers that will be published over time, what really gives a more up-to-date picture of the ongoing situation are infection numbers based on registration (rather than reporting) date. It is possible to combine forecasting and nowcasting to predict cases by registration date, thus setting the real number of infections being discovered during a specific set of dates as target variable. This process, that we call \textsl{forenowcasting}, is equivalent to filling the blue square on the bottom of Figure \ref{fig:sketch}. This is done by computing forecasts as described in the previous subsection, and then performing nowcasting on the forecasted numbers. We also obtain uncertainty estimates in an analogous way as for forecasts and nowcasts. 

%Quantitative measures of the quality of the predictions over time were shown in Section \ref{sec:evaluation}. In the following section, we compare, in exlempary fashion, predictions obtained for two specific dates with their realized counterparts, observed . 

\subsection{Retrospective Surveillance}

 %for nowcasting, forecasting and forenowcasting, which we show in Figure \ref{fig:predcitions}.

\begin{figure}[]
	\centering
\includegraphics[width = 0.4 \textwidth]{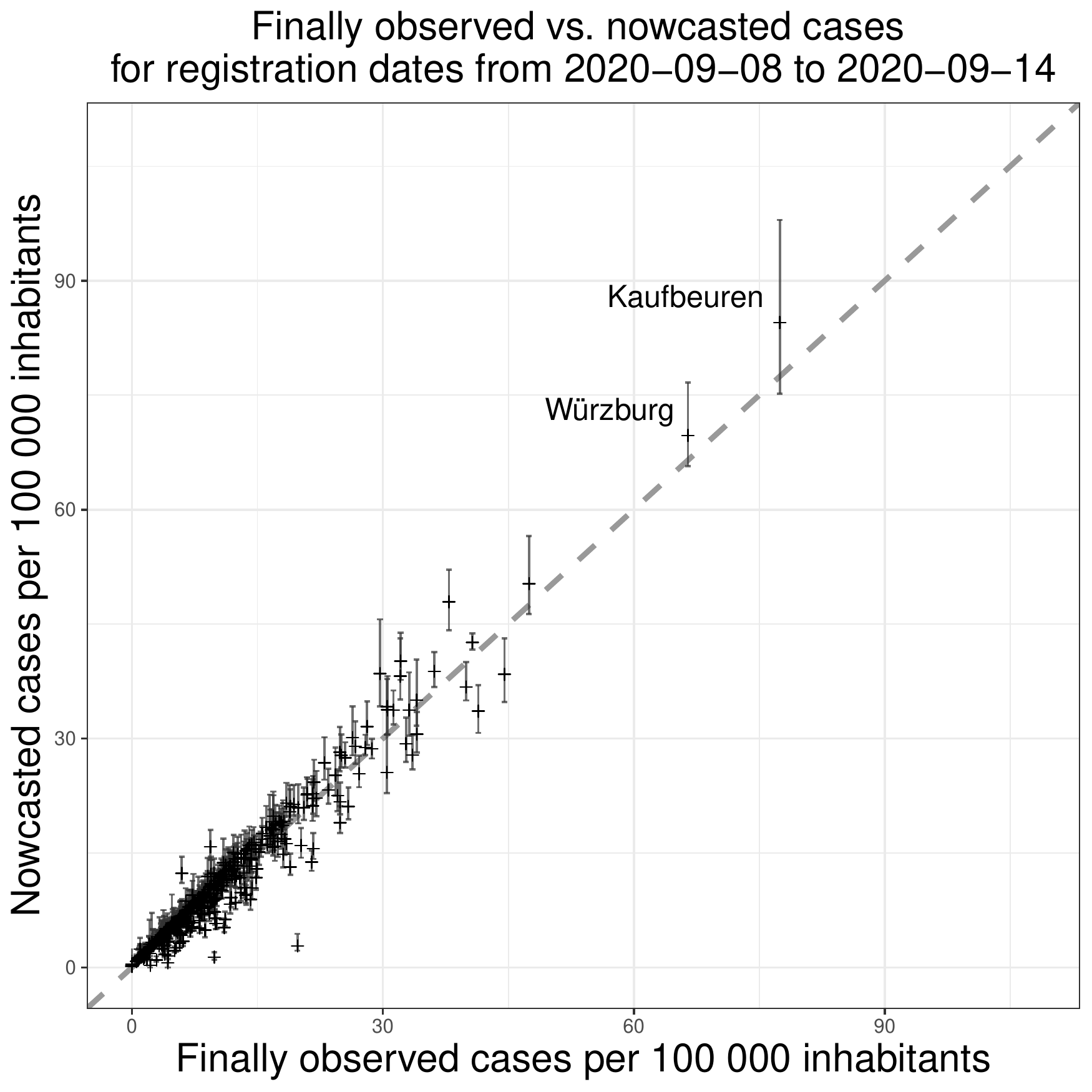}
\includegraphics[width = 0.4 \textwidth]{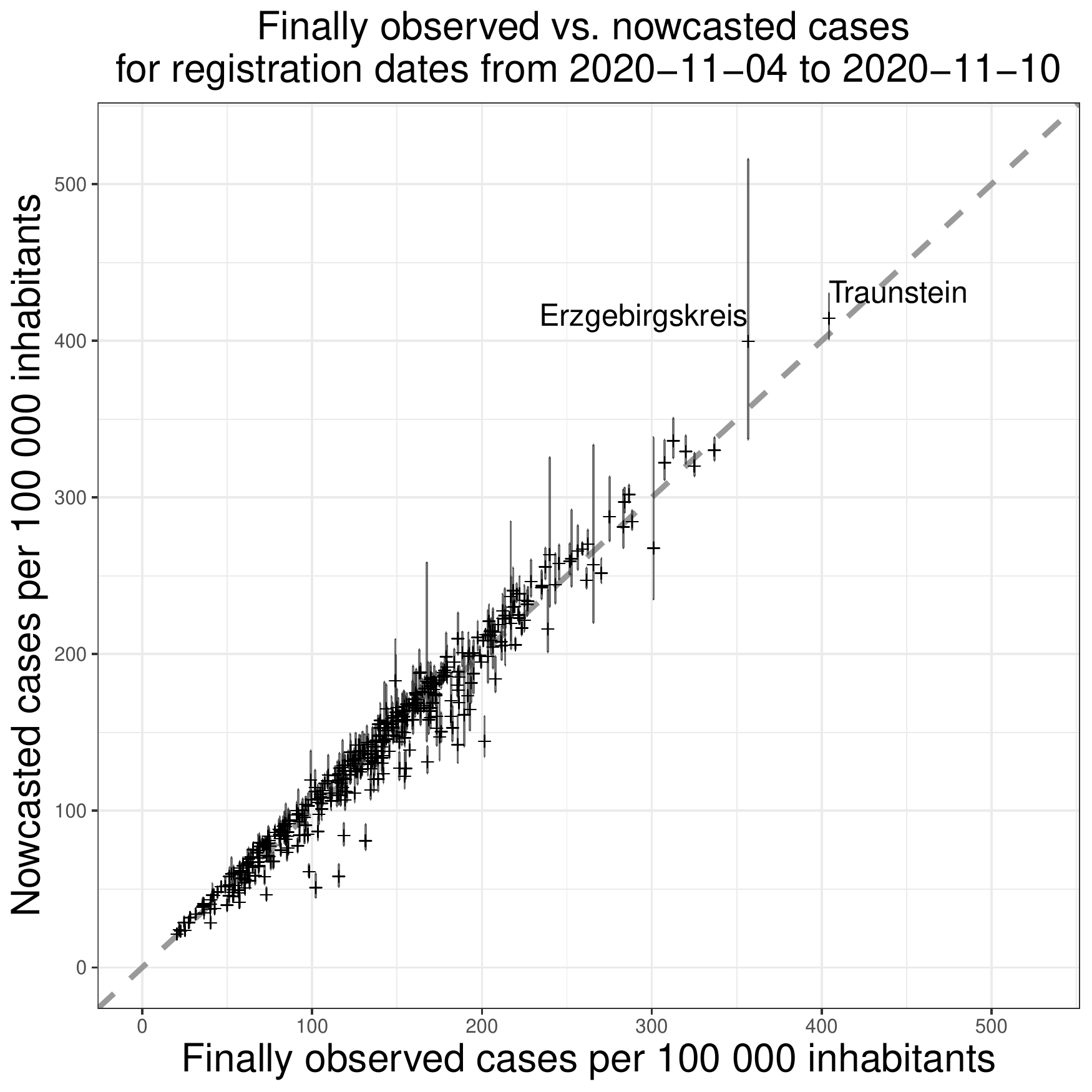} \\
\includegraphics[width = 0.4 \textwidth]{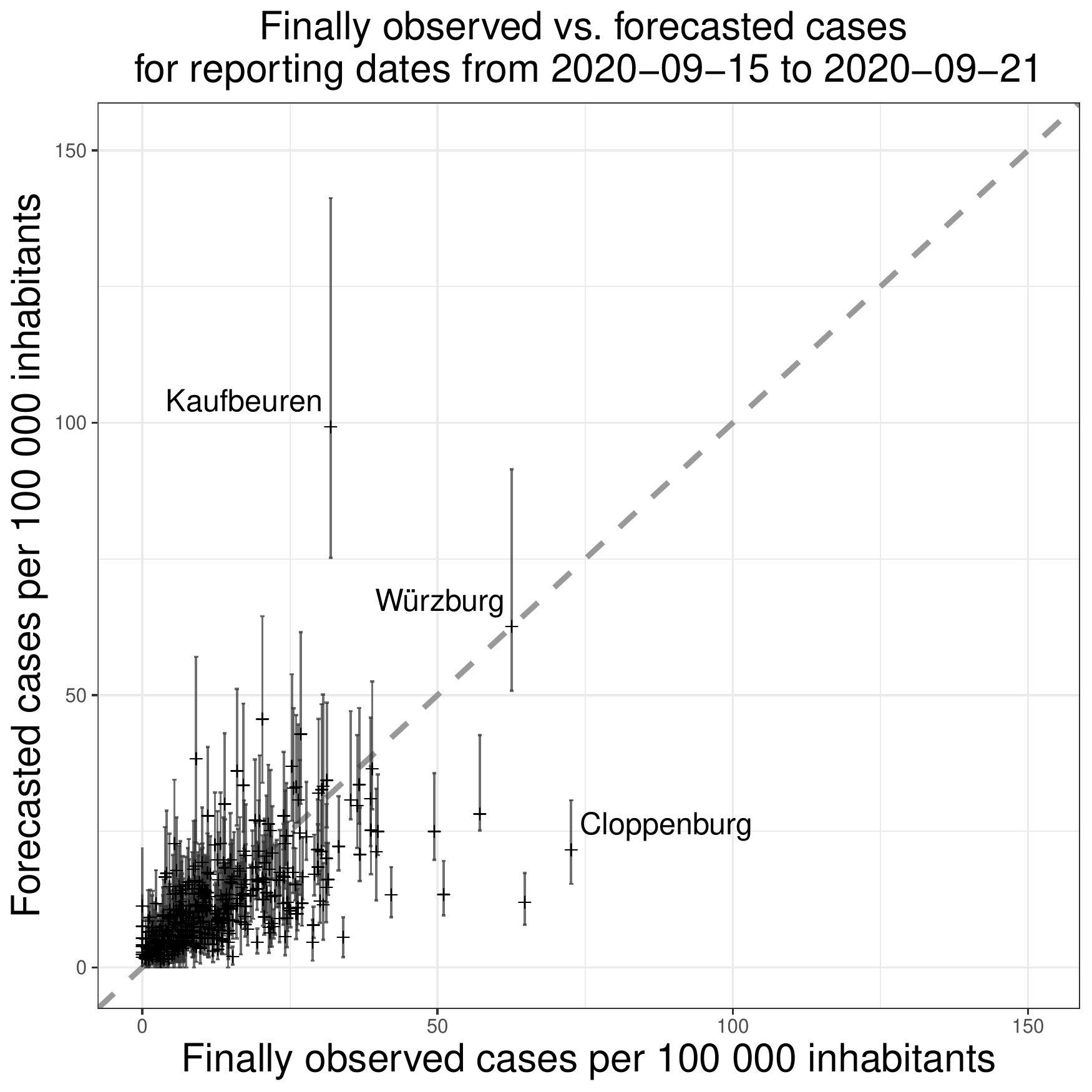} 
\includegraphics[width = 0.4 \textwidth]{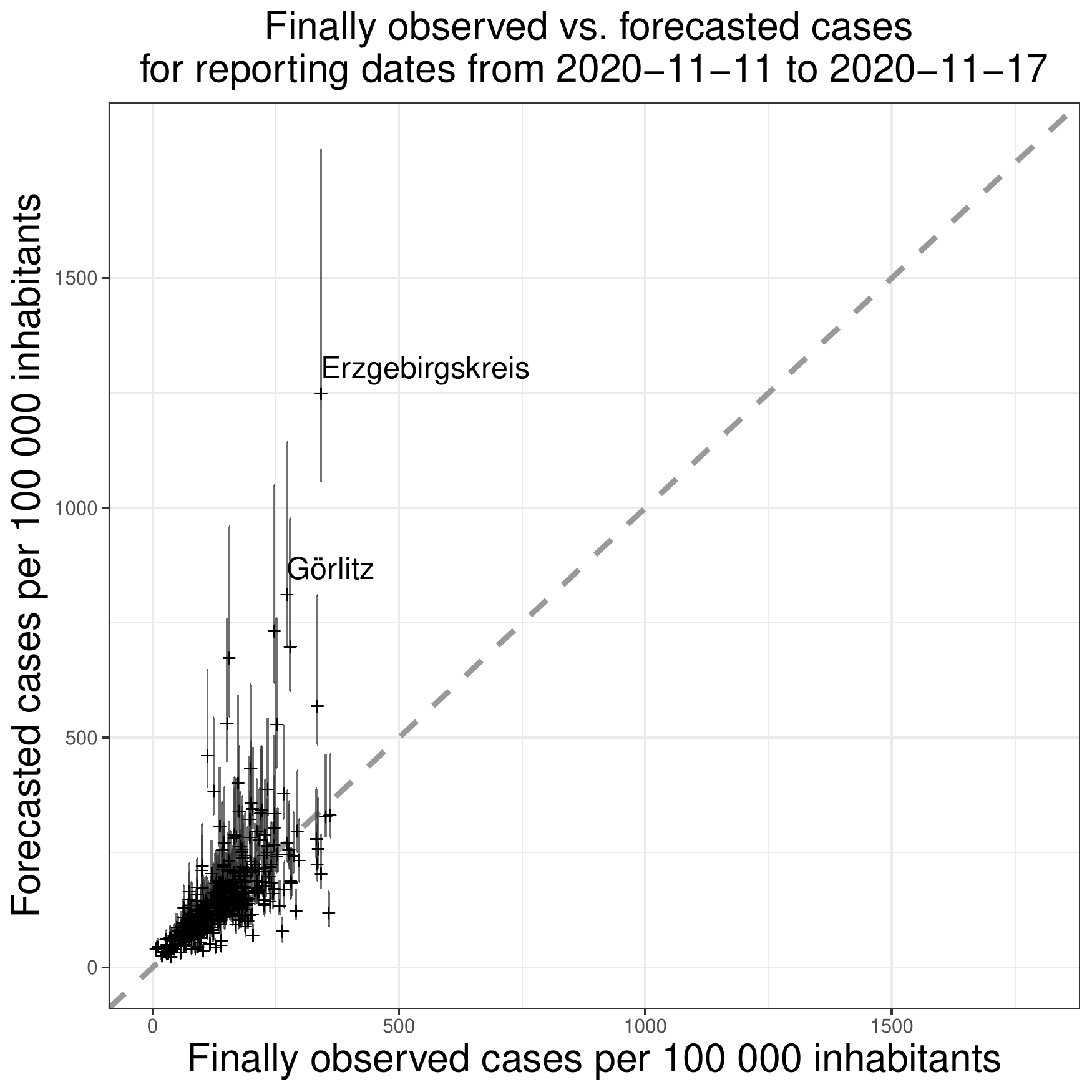} \\
\includegraphics[width = 0.4 \textwidth]{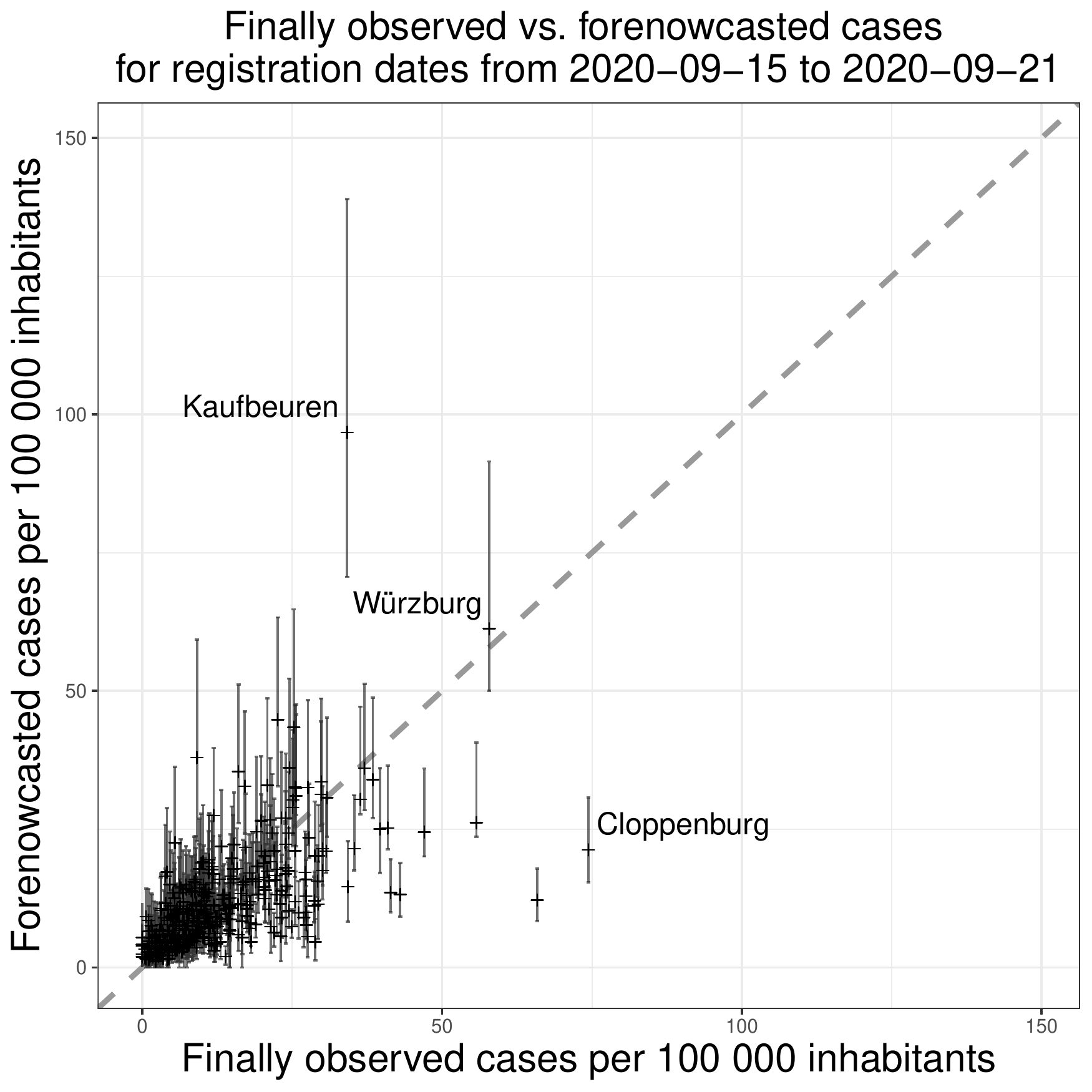} 
\includegraphics[width = 0.4 \textwidth]{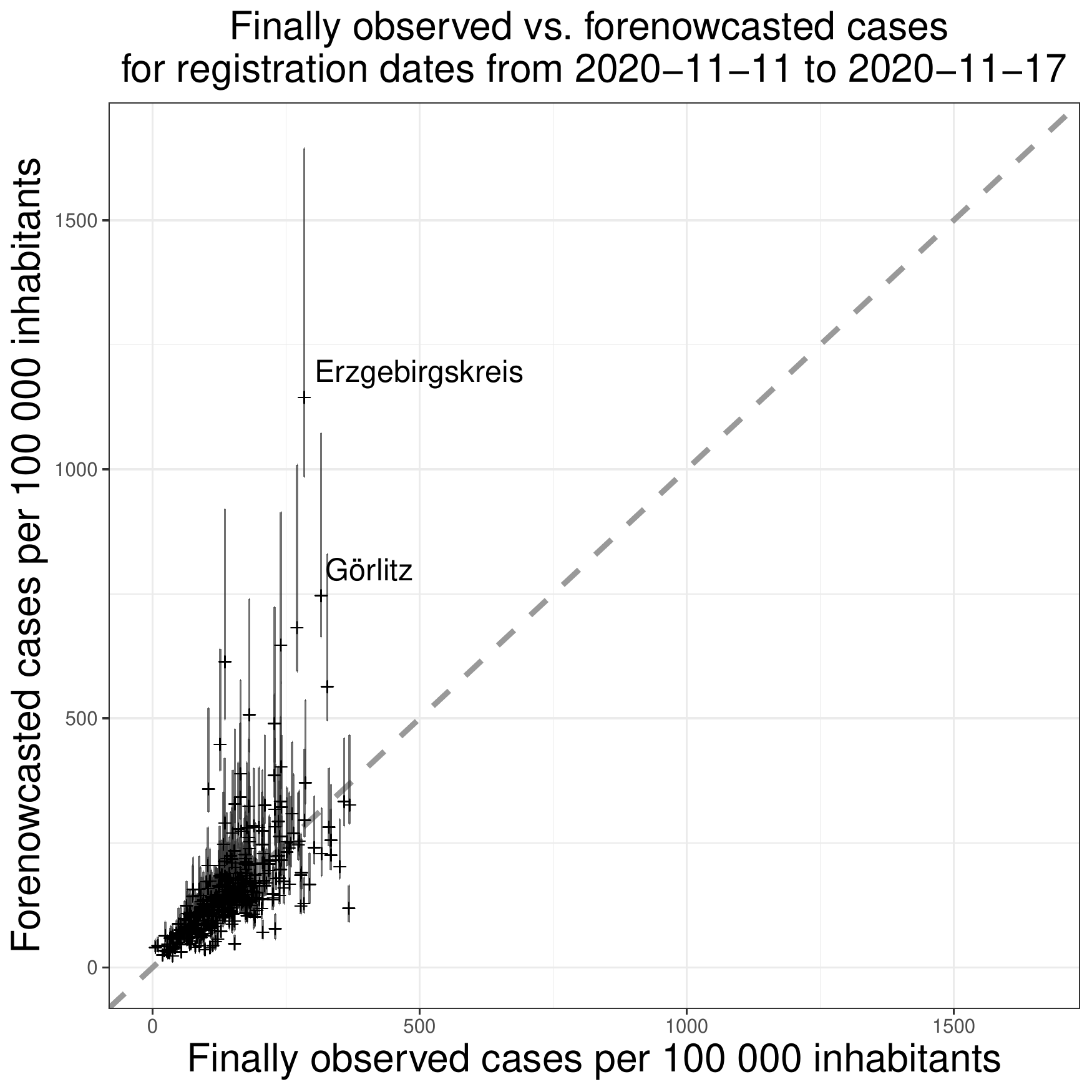} 
\caption {Nowcasts (top), forecasts (middle) and forenowcasts (bottom) of cumulated infections over a week, cumulated by district, plotted against values observed \textit{a posteriori}. The model is fitted with data available on the dates September 15, 2020 (left) and November 11, 2020 (right). Vertical lines represent prediction intervals computed at the 90\% level. }
\label{fig:predcitions}
\end{figure}

It is also possible to utilize the proposed model as a surveillance tool retrospectively. After a certain period of time has passed from the day of analysis, we are able to compare predictions with infections observed in the corresponding period. If the predictions are aggregated on a weekly basis and we keep the maximum delay set as $d_{max}=7$, the waiting time to observe realized infection numbers will be equal to seven days for nowcasts and forecasts and fourteen days for forenowcasts. Figure \ref{fig:predcitions} shows predictions of all three kinds and corresponding infections observed \textsl{a posteriori} for two exemplary days of analysis, namely September 15 (left hand panel) and November 11, 2020 (right hand panel).

%Setting $d_{max}=7$ we are able to compare the results of the nowcasts with realized infections a week after the analysis. The top panel of Figure \ref{fig:predcitions} shows precisely that comparison, with cumulated nowcasted cases for the week from September 8 to September 14, 2020 using data available on September 15, 2020 for each district on the $y$-axis and the corresponding realized cases, observed in full a week later, on the $x$-axis. Vertical lines indicate prediction intervals at the 90\% level. From the plot we can observe how the nowcasts are able to capture most of the residual variability in the data, including that for the two most affected districts in the country at the time of the analysis, namely Kaufbeuren and Würzburg. %The biggest error in the nowcast is observed for the district of Oldenburg, in which apparently an anomalously large number of cases were reported with delay. 

From the plots we can observe how nowcasts tend to be, in general, quite precise, as already seen from Figure \ref{fig:errors}.
We can also immediately notice how performance is very different for the two dates, especially for forecasts and forenowcasts: We see that the predictions for September 15th are relatively precise and unbiased, while for November 11th we observe quite a strong tendency towards overprediction. This is because, while the first date belongs to a period in which the pandemic was relatively stable in Germany, the second one lies in the heart of the second wave. The latter was immediately successive to the sudden increase in new infections in October and to the consequent implementation of social distancing measures, the so-called ``lockdown light'', from the beginning of November. However, our model does not include anything regarding exogenous governmental interventions and general changes in population behavior. This means that the predictions are to be interpreted assuming that everything else stays the same as in the three weeks used to fit the model, leading to overprediction for areas in which measures are indeed imposed, and possibly underpredictions after those measures are softened or lifted. While this is certainly a limitation of our approach, it can on the other hand also be seen as a feature of the model, which in a sense provides potential future "counterfactual" scenarios in which no action was taken by decision makers. This can thus be used to try to quantify the effect of social distancing policies and interventions, in specific districts as well as at a broader level. % An example of this is the prediction for the district of Gütersloh for the model fitted using data available on September 15, 2020, just a week later than the week we displayed. While the model predicted more than 1000 weekly cases per 100.000 inhabitants, a truly disastrous proposition, the realized number of infections in that week was just around 125 cases per 100.000, presumably thanks in large part to non-pharmaceutical interventions (NPI). 

This also applies in the case of sudden outbreaks: If, in a specific district, a rapid spike in cases is observed which was not yet known to health authorities at the time of the analysis, the model will naturally underpredict infection numbers in that district. %After all, of course, the model predicts future infections with data from the past, and is not clairvoyant.  %An example for this are the September predictions for the district of Cloppenburg, Lower Saxony, in which an outbreak occurred during the forecasted period.
Severe underpredictions observed \textsl{a posteriori} can also be used as an indicator for ``true'' outbreaks, revealing if they were explainable by past data or not. This ``counterfactual'' use of our model can thus be seen as an additional feature, which becomes available in retrospect, to measure the effect of NPIs (Non-Pharmaceutical Interventions) and to assess the nature of outbreaks.

\section{Discussion}
\label{sec:discussion}

We proposed a numerically stable tool to nowcast and forecast COVID-19 cases reported with delay. This allows to perform surveillance by gender and age group at the regional level, providing an up-to-date and detailed picture of the pandemic, as well as giving insight in the dynamics of the near future. Our model can be used for computing inherently dynamic index measures, such as the 7-days incidence, both at the regional and national level, and it can also aid governments in the implementation of more targeted area- and population-specific containment strategies. However, as previously mentioned, this approach does not come without limitations, which we also want to address.

The number of detected cases greatly depends on local testing strategies and capacities. This implies that comparisons between different states or regions are not straightforward. As our model makes use of reported infections, direct comparisons should  be limited to areas for which it is reasonable to assume that testing has been carried out in a similar manner.   

Another important thing to note is that our model only addresses the delay in reporting from local to national health authorities, and not the time that occurs between each test and its (positive) result. This would be useful for our application as it would give an even more up-to date picture of the current situation, but it is not pursued due to a lack of data.

An eminent  limitation of our approach is the inability to capture new outbreaks related to specific phenomena that are not yet known to the health authorities. An example of this would be the outbreaks in slaughterhouses which happened during the summer in Coesfeld and Gütersloh, North-Rhine-Westphalia. After all, of course, the model predicts future infections with data from the past, and is not clairvoyant. As previously discussed, severe underpredictions observed \textsl{a posteriori} can also be used in retrospect as an indicator for outbreaks that are localized and not explainable by past data, while overpredictions can signal and quantify the effectiveness of social distancing measures. 

%This last point is motivated by the fact that our model does not include anything regarding exogenous social distancing measures. This means that the predictions are to be interpreted assuming that everything else stays the same, leading to overprediction for areas in which measures are indeed imposed. On the other hand, this can also be seen as a feature of the model, which in a sense provides potential future "counterfactual" scenarios in which no action is taken by decision makers. %A prime example of this is the prediction for the previously mentioned district of Gütersloh for the model fitted using data available on July 6, 2020, just a week later than the week we displayed. While the model predicted more than 1000 weekly cases per 100.000 inhabitants, a truly disastrous proposition, the realized number of infections in that week was just around 125 cases per 100.000, presumably thanks in large part to non-pharmaceutical interventions (NPI). 

Taking into account the previously mentioned limitations, the model is able to capture a good chunk of the variability that is present.
The methodology that we employed is quite general, and, if suitable data is available, can easily be adapted to other countries as well. 
Moreover, we only employed standard tools for software implementation, and this makes adapting and enriching the model, e.g. with more covariates, relatively straightforward. We here  did not include more covariates which would be available for the special case of Germany (e.g. district-wise deprivation indexes, see \citealp{Maier2017}), because we want our modeling exercise to be flexible and easy to apply to any country in which delays in reporting happen, with basic data being available. Nonetheless, it would most certainly be fruitful, at least from a predictive accuracy perspective, to include more covariates which are available for specific cases in the model.

We complete our discussion by emphasizing that the proposed methodology is flexible and applicable to any data constellation in which reporting delay plays a role. In other words, one can easily adopt the proposed model to any guillotine blade-like data structures, i.e.\  data where $t_i$ denotes the time point of an event and $d_i$ the delay with which the event is reported. 
Moreover, our approach can not only be applied to correct for the delay between registration of an event and its reporting, but also, for example, to bridge the delay between disease onset and registration of its positive test result.  
Data in guillotine blade-like form also occur in areas beyond epidemiology, e.g.\  when cases of unemployment are reported from regional offices to a central state register. The generality of the data structure supports the proposed modeling approach, where corrections for the missing data structure are directly incorporated in the model.  
In particular, however, the modeling exercise exhibits promising performance for COVID-19 infections, and may therefore be incorporated into a general surveillance tool to assist health authorities and policymakers in their efforts to contain the spread.

\clearpage

\bibliographystyle{chicago}

\bibliography{references}

\clearpage

\end{document}